\documentclass[aps,prl, floatfix, reprint, superscriptaddress]{revtex4-1}

\usepackage{graphicx}
\usepackage{color}
\usepackage[normalem]{ulem}
\usepackage{hyperref}
\usepackage{amsmath}
\usepackage{lipsum}
\usepackage{xcolor}

\begin{abstract}
Time-dependent nonlinear media, such as rapidly generated plasmas produced via laser ionization of gases, can increase the energy of individual laser photons and generate tunable high-order harmonic pulses. This phenomenon, known as photon acceleration, has traditionally required extreme-intensity laser pulses and macroscopic propagation lengths. Here, we report on a novel nonlinear material---an ultrathin semiconductor metasurface---that exhibits efficient photon acceleration at low intensities. We observe a signature nonlinear manifestation of photon acceleration: third-harmonic generation of near-infrared photons with tunable frequencies reaching up to $\approx3.1\omega$. A simple time-dependent coupled-mode theory, found to be in good agreement with experimental results, is utilized to predict a new path towards nonlinear radiation sources that combine resonant upconversion with broadband operation.
\end{abstract}

\begin{document}

\title{Nonlinear manifestations of photon acceleration in time-dependent metasurfaces: tunable broadband harmonics generation}

\author{Maxim~R.~Shcherbakov}
\email{mrs356@cornell.edu}
\affiliation{School of Applied and Engineering Physics, Cornell University, Ithaca, NY 14853, USA}

\author{Kevin~Werner}
\affiliation{Department of Physics, The Ohio State University, Columbus, OH 43210, USA}

\author{Zhiyuan~Fan}
\affiliation{School of Applied and Engineering Physics, Cornell University, Ithaca, NY 14853, USA}

\author{Noah~Talisa}
\affiliation{Department of Physics, The Ohio State University, Columbus, OH 43210, USA}

\author{Enam Chowdhury}
\affiliation{Department of Physics, The Ohio State University, Columbus, OH 43210, USA}

\author{Gennady~Shvets}
\email{gshvets@cornell.edu}
\affiliation{School of Applied and Engineering Physics, Cornell University, Ithaca, NY 14853, USA}

\date{\today}

\maketitle

Before the demonstration of the first laser by Theodore Maiman, light propagation was widely considered to be a linear process, with the photons not expected to interact with each other. This simple understanding of light-matter interactions was overturned in the early 1960's in second harmonic generation experiments by Franken \textit{et al.} \cite{Franken1961}. From this demonstration of the merger between two photons into a photon with doubled energy, the nonlinear optics was born. Subsequent realizations of the third \cite{Maker1986} and higher-order \cite{Burnett1977} harmonics enabled efficient light sources \cite{Yamada1993}, high-resolution microscopy \cite{Campagnola2003,Debarre2006}, and produced some of the most sensitive optical characterization techniques \cite{Heinz1982,Shen1989,Shurik2011}.

However, fundamental effects limit the efficiency and spectral range of the canonical nonlinear processes. Mainly, the very nature of the standard $n$-photon processes (where $n \geq 2$ is an integer) dictates that a narrow-band laser pulse centered at the frequency $\omega_L$ with the spectral width $\Delta \omega_L$ cannot produce upconverted photons with the frequencies $\Omega_n$ outside of the $|\Omega_n - n\omega_L|\lesssim \sqrt{n} \Delta \omega_L$ spectral interval. While using high-finesse optical cavities or other resonant structures can enhance the efficiency of harmonics generation, they even further limit the spectral range of the nonlinearly generated photons. Thus, a fundamental challenge is to find a nonlinear optical process that enables efficient nonlinear frequency conversion without sacrificing the spectral bandwidth. In this Article, we propose and experimentally realize one such process: upconversion of mid-infrared (MIR) light undergoing rapid blue-shifting---also known as photon acceleration \cite{Mendonca2000}---in a resonant nonlinear metasurface.

The concept of photon acceleration (PA) was originally introduced in gaseous plasmas \cite{Yablonovitch1974,Wilks1989} as a process of frequency conversion that occurs when electromagnetic waves propagate in a medium with a time-dependent refractive index \cite{Felsen1970}. A reduction of the refractive index via free carrier (FC) generation results in a measurable blue-shifting regardless of whether the FCs were produced by the radiation itself~\cite{Yablonovitch1974,Wood1991} or by an auxiliary electromagnetic pulse~\cite{Savage1992}, as well as in the broadening of the spectrum \cite{Yablonovitch1973}, which was demonstrated for harmonics generation as well \cite{Siders1996}. PA in a solid (e.g., semiconductor) medium can be achieved at much lower laser intensities than in a gas because of the ease of FC generation \cite{Turchinovich2012,Blanco2014}, and can be further enhanced in high quality factor (high-$Q$) optical cavities. For example, loading photons into a ring microcavity \cite{Preble2007} or a photonic crystal cavity \cite{Tanabe2009} and subsequently generating FCs by an external pump while the photons are still in the cavity resulted in continuous near-infrared wavelength shifts of up to several nanometers \cite{Preble2007,Dong2008}. However, no nonlinear manifestations were observed, primarily because of high sensitivity of high-Q resonators to high-intensity near-infrared light. Therefore, new photon-accelerating platforms based on free-space light coupling are needed.

Recently, a new paradigm of regularly nanostructured surfaces---metasurfaces~\cite{Holloway2012,Kildishev2013,Yu2014a}---has been established for ultra-thin nonlinear and active materials \cite{Li2017,Krasnok2017}. While metasurfaces share with optical cavities the attractive properties of high spectral selectivity and strong field concentration, their important feature is the strong coupling to free-space beams. A variety of metasurface designs have been implemented for applications as diverse as wave front manipulation~(in both linear \cite{Yu2011} and nonlinear \cite{Li2015} regimes), rapid amplitude and phase modulation~\cite{Emani2012,Yao2013,Dabidian2015,Dabidian2016}, as well as efficient harmonics generation \cite{Klein2006,Lee2014,Vampa2017} and all-optical modulation \cite{Wurtz2011,Guo2016,Shcherbakov2017}. Of particular interest are semiconductor-based metasurfaces that utilize strong, geometry-dependent Mie-type localized modes \cite{Kuznetsov2016a} with high $Q$-factors~\cite{Wu2014,Yang2014a}. They have already shown record-breaking nonlinear-optical performance on the nanoscale
\cite{Shcherbakov2014b,Liu2016c,Grinblat2016,Makarov2017}, making them an attractive platform for observing PA and tunable and broadband optical harmonics.

\begin{figure*}
\includegraphics[width=1\textwidth]{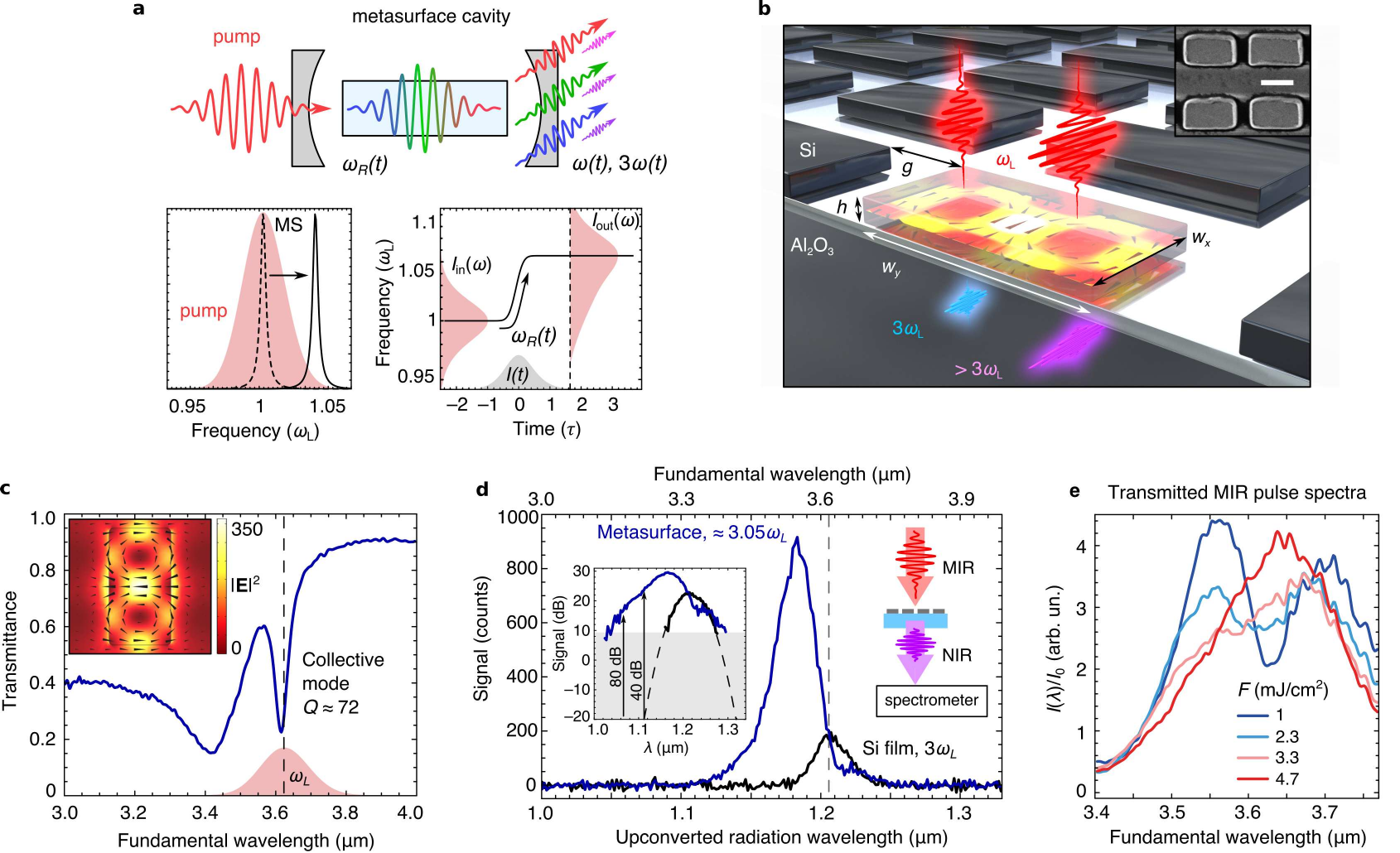}
\caption{\label{fig1} \textbf{Self-induced blue-shift of harmonics generation from a nonlinear photon accelerating semiconductor infrared metasurface (PASIM).}
\textbf{a}, The concept of  blue-shifted harmonics generation: MIR photons are trapped by the metasurface cavity, blue-shifted by the rapid refractive index variation due to FC generation inside the meta-surface, and then nonlinearly upconverted to NIR photons via the standard THG process. The blue-shifted MIR and NIR photons then leave the metasurface, and their spectra are detected in transmission. \textbf{b}, Schematic of the sample and the MIR beam setup. In experiments, the following dimensions of the sample were used: $w_x = 0.87$~$\mu$m, $w_y = 1.54$~$\mu$m, $g = 400$~nm, and $h=600$~nm. Inset: a close-up scanning electron micrograph of the metasurface elements; scale bar: $1~\mu$m.
\textbf{c}, Transmittance spectra of the metasurface measured using FTIR spectroscopy. Dashed line: central wavelength $\lambda_L$ of the MIR fs pulses. Inset: simulated electric field $\mathbf E$ (arrows) and intensity $|\mathbf E|^2$ (color) inside a metasurface unit cell for the incident light at $\lambda=\lambda_L$.
\textbf{d}, Experimentally measured THG in the metasurface (blue curve) and in an unstructured Si film (black curve) of the same thickness $h$. Both spectra are measured at the same MIR fluence of $F=2.3$~mJ/cm$^2$ (peak intensity $I=11$~GW/cm$^2$). A notable blue-shift of the upconverison peak from the metasurface with respect to the expected THG is a result of the plasma-induced acceleration of MIR photons. 
Inset: the same data plotted on the logarithmic scale. Grey region: the noise floor. The THG from the metasurface extends to $\lambda=1.06$~$\mu$m, with intensities eight orders of magnitude higher than projected from the unstructured film (Gaussian fit given in dashed line): a clear indication of the emergence of the photons that are not present in the incident spectrum.
\textbf{e}, Spectra of MIR pulses transmitted through the metasurface for different input fluences revealing significant pulse self-modulation.}
\end{figure*}

Here, we design and experimentally realise a photon accelerating semiconductor infrared metasurface (PASIM) that undergoes rapid refractive index changes due to highly-nonlinear photoinduced generation of free carriers  (FCs) in silicon by a MIR pulse. The main idea of photon acceleration in a PASIM is given in Fig.~\ref{fig1}(a). Briefly, MIR photons interact with, and get trapped by, the metasurface. As FCs are generated by four-photon absorption (4PA), the resonant frequency of the metasurface blue-shifts, and the frequency of the trapped photons follows. Accelerated MIR photons then upconvert via the standard $\hat\chi^{(3)}$ nonlinear process
, resulting in the observed blue-shifting of the THG. The PASIM is designed to have a high-$Q$ resonance at $\lambda_R \equiv 2\pi c/\omega_R = 3.62$~$\mu$m that enables efficient four-photon FC generation at modest pulse intensities.
This enables us to clearly observe the effect of PA on harmonics generation, with the peaks of the upconverted radiation appearing at frequencies of up to $\Omega_n\approx 3.1 \omega_L$, where $\omega_L$ is the central frequency of the MIR pulses chosen at $\omega_L \approx \omega_R$ to maximize the effect. 
Moreover, we detect anomalous levels of nonlinearly generated signal with frequencies of up to $\approx3.4 \omega_L$ in the wings of the THG spectra, corresponding to the spectral density enhancement of $\approx10^8$ over the projected signal from an unstructured film. Such enhancement can only be explained by the frequency boost of the spectrally-dense population of photons with initial frequencies around $\omega_L$, and their subsequent nonlinear conversion. An intuitive coupled-mode theory model with time-dependent eigenfrequency $\omega_R(t)$ and damping factor $\gamma_R(t) \equiv \omega_R(t)/Q_R(t)$ (where $Q_R(t)$ is a time-dependent quality factor of the metasurface) accurately captures most features of the experimental data and provides further insights into PA efficiency improvements, thus paving the way to future applications utilizing non-perturbative nonlinear nanophotonics.

\section{Results}
\textbf{Metasurface design and fabrication.} The metasurface was engineered to enhance the local fields, which is crucial for the efficiency of the nonlinear photon conversion and multiphoton absorption. In our design of the metasurface, we make use of high-$Q$ collective resonances common in regular arrays of semiconductor particles \cite{Wu2014,Yang2014a,Shalaev2015}. The specific implementation of the PASIM comprised of nearly-touching rectangular Si nanoantennas is shown in Fig.~\ref{fig1}b. The quality factor of the resonance---and, hence, the local field enhancement---is controlled by the gap $g$ between the rectangles (see the Supplementary Note~1 for ultrahigh $Q$s of up to $\approx 10^4$). For photon acceleration of broadband femtosecond laser pules, the $Q$-factor of the PASIM was designed to be moderate: $Q\approx 72$, as experimentally confirmed by Fourier-transform infrared (FTIR) spectroscopy measurements using collimated beams~\cite{Wu2014}, as shown in Fig.~\ref{fig1}c. Using full-wave  simulations, we determine the local MIR field enhancement to be up to $|\mathbf{E_{\mbox{\tiny loc}}}/\mathbf{E}_0|^2=350$, where $\mathbf{E}_0$ is the amplitude of the incident electric field polarized along the short dimension of the rectangle. The sharpness of the resonance is attributed to its very small dipole strength due to the near-complete cancellation between positive and negative polarization densities inside the nanoantennas~\cite{Neuner2013} as shown in the inset of Fig.~\ref{fig1}c. The samples were fabricated according to a standard procedure described in Methods and Supplementary Note~2.

\textbf{Upconversion spectroscopy.}
Intense ultra-short MIR laser pulses with a variable non-destructive fluence in the $1 < F < 6$~mJ/cm$^2$ range, with central wavelength $\lambda_L=3.62$~$\mu$m and time duration $\tau_L \approx 200\pm30$~fs (spectral FWHM $\Delta \lambda_L \approx 150\pm50$~nm) were focused onto the PASIM to $5 <I <30$~GW/cm$^2$ peak intensity (see Supplementary Note~3 for the details of the optical setup).
 The advantage of operating the PASIM in the MIR regime is that the refractive index change scales as $\Delta n(t) \propto - N_{\rm FC}(t)\lambda^2$, which will be crucial for PA; here, $N_{\rm FC}(t)$ is the time-dependent FC density. Note that we neglect Kerr and thermal additions to the refractive index; corresponding discussion and estimates are provided in Supplementary Note 4. Kerr effect may play an important role in frequency conversion processes and may present an interesting avenue for further research in metasurfaces \cite{Drozdov2012,Buyanovskaya2017}.

The third-order nonlinear polarization, which is responsible for the third harmonic generation (THG), can be expressed in the perturbative regime as follows:
\begin{equation}\label{eq:thg}
\mathbf{P}^{(3)}(3\omega)=\varepsilon_0\hat{\chi}^{(3)}(3\omega= \omega+\omega+\omega)\vdots\mathbf{E}(\omega)\mathbf{E}(\omega)\mathbf{E}(\omega),
\end{equation}
where $\hat{\chi}^{(3)}(3\omega= \omega+\omega+\omega)$ is the third-order nonlinear susceptibility tensor of silicon, and $\mathbf{E}(\omega)$ is the local electric field strength at the pump frequency $\omega$. Although the resonant local field is confined to a very small volume of each nanoantenna, the cubic dependence of $\mathbf{P}^{(3)}$ on the local field enables considerable THG boost in the metasurface as compared to the unpatterned silicon film. This is indeed experimentally observed in the transmitted THG spectrum plotted in Fig.~\ref{fig1}d for the metasurface (blue curve) versus the unpatterned film of the same thickness (black curve) cases: an order of magnitude spectrally-integrated THG enhancement is  provided by the metasurface, and a maximum conversion efficiency estimated at around $10^{-9}$. While similar magnitudes of THG enhancement and conversion efficiencies have been observed in the past \cite{Shorokhov2016a}, much more revealing is the {\it spectrum} of the THG that has never been collected from rapidly-changing metasurfaces, and which reveals several unambiguous signatures of the PA process.

We observe three features of the THG spectra that have not been previously experimentally observed or theoretically predicted for a solid-state medium. All three indicate that the THG takes place under non-perturbative conditions, when the optical properties of the metasurface are strongly modified while the laser pulse is interacting with it. {\it First,} the spectral peak is strongly blue-shifted with respect to its unperturbed $3\omega_L$ (corresponding to $\lambda_{\rm TH}^{(0)} \approx 1.207$~$\mu$m) spectral position measured with an unpatterned film, and with respect to the tripled frequency of the PASIM resonance at $3\omega_R$. {\it Second,} because the spectral shift of the peak to $\approx 3.05\omega_L$ is larger than the full widths at half maxima (FWHMs) of the unperturbed THG spectrum and of the metasurface resonance, there exists a high-frequency spectral region of the THG ($\lambda_{\rm TH} < 1.17$~$\mu$m, on the blue side of the spectral peak) where the THG signal from the metasurface is more than two orders of magnitude higher than that from the Si film. This is a signature of the photons accelerated from the spectral peak of the incident pulse outside of its spectral width. 
In the inset of Fig.~\ref{fig1}d, we show that signal with the wavelengths as short as $\lambda_{\rm TH} \sim 1.06 \mu$m (i.e. more than $100$~nm on the blue side of the highest energy signal detected from the unstructured film) can be detected. For example, for two specific wavelengths, $\lambda_{\rm TH} = 1.12$~$\mu$m and $\lambda_{\rm TH} = 1.06$~$\mu$m, the measured THG intensities from the metasurface are, respectively, four and eight orders of magnitude stronger than the corresponding projections calculated from the THG spectrum that was collected from the unpatterned film (dashed line). Note that it is crucial to differentiate between the generation of new photon frequencies, as it happens in the case of the acceleration of photons trapped by a high-$Q$ resonator, and the reshaping of the laser spectrum due to the nonlinear modification of low-$Q$ optical resonances \cite{Zhang2011,Zhang2012}.
{\it Finally,} the THG spectrum reveals another counter-intuitive property of a PASIM: it is possible to resonantly enhance a nonlinear process (THG) without sacrificing the spectral bandwidth. All three features are related to the emergence of the new, higher energy photons, and can be attributed to photon acceleration due to the dynamic multiphoton FC generation.
The transmitted MIR spectra shown in Fig.~\ref{fig1}e reveal strong power dependence, thus providing further confirmation of the non-perturbative nature of the PA. No such dependence on the MIR fluence $F$ was observed in bulk silicon (Supplementary Note~4), thus validating the key role of the resonantly excited hot spots that enable FC generation through 4PA.

\begin{figure}[t!]
\includegraphics[width=\columnwidth]{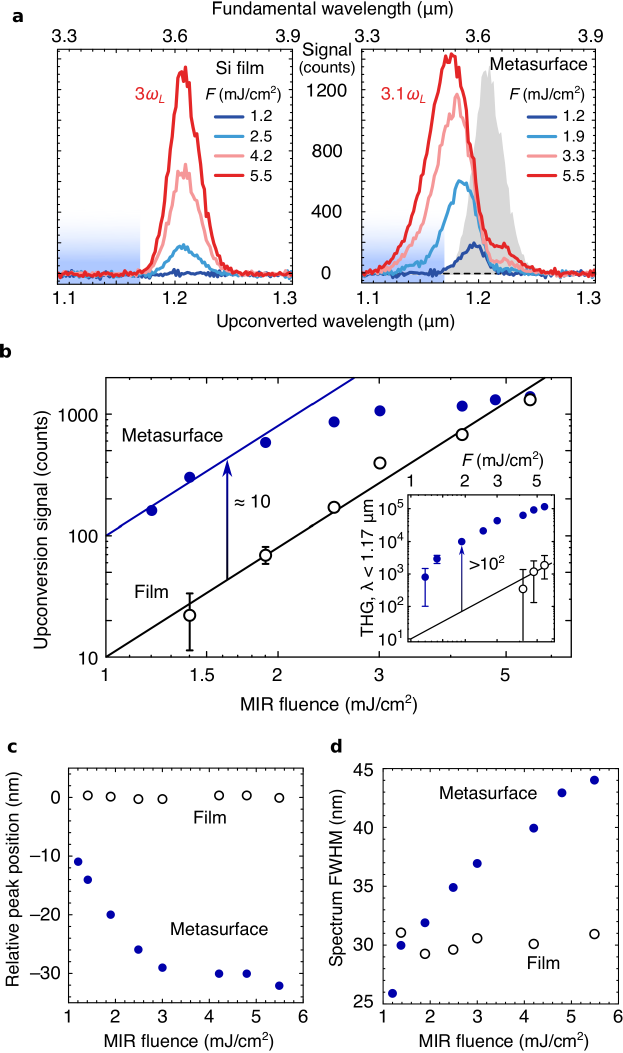}
\caption{\label{fig2} \textbf{Spectral characteristics of blue-shifted harmonics in PASIMs.} \textbf{a},
The upconversion NIR spectra measured for various input fluences for the unpatterned Si film (left) and for the metasurface (right). Shaded grey area in the right panel: same as in the left panel at the highest fluence $F_{\rm max} = 5.5$~mJ/cm$^2$. Shaded blue areas indicate the integration range of $\lambda_{\rm TH}<1.17$~$\mu$m.  \textbf{b}, Spectrally integrated THG as a function of the MIR fluence the film (open circles) and metasurface (closed circles), log--log scales. Lines denote the guide-to-the-eye $I_{\rm TH} \propto F^3$ dependence. Inset: THG spectrum integrated over $1.1$~$\mu\mbox{m}<\lambda_{\rm TH}< 1.17$~$\mu$m as a function of  $F$ for the unstructured film (open circles) and the metasurface (filled circles). The THG spectrum from the PASIM reveals self-induced blue-shifting by $\approx30$~nm (\textbf{c}) and broadening by $\approx 50\%$ (\textbf{d}) as a function of $F$; both effects are negligible for the unpatterned film.}
\end{figure}

To quantify the combined process that manifests as blue-shifted harmonics generation, we measured the NIR spectra as a function of the incident MIR fluence $F$ (Fig.~\ref{fig2}c), and compared them with the corresponding spectra generated in the unpatterned Si film (Fig.~\ref{fig2}b). 
We observe that the spectral peak and width of the THG light generated in the PASIM can be controlled by incident fluence. The central THG  wavelength can be blue-shifted by more than $30~$nm, as shown in Fig.~\ref{fig2}e, enabling the THG with center frequencies of up to $\approx 3.1\omega_L$. In contrast with the common belief that the resonant enhancement of the THG must be accompanied by spectral narrowing, our results plotted in Fig.~\ref{fig2}f indicate the opposite. The upconverted signal has a spectrum that is up to $50\%$ broader than that from the unstructured film, which is a clear fingerprint of PA. We therefore conclude that the perturbative approach expressed by Eq.(\ref{eq:thg}) fails, suggesting the need for a more accurate model of harmonic upconversion.

\begin{figure*}
\includegraphics[width=0.9\textwidth]{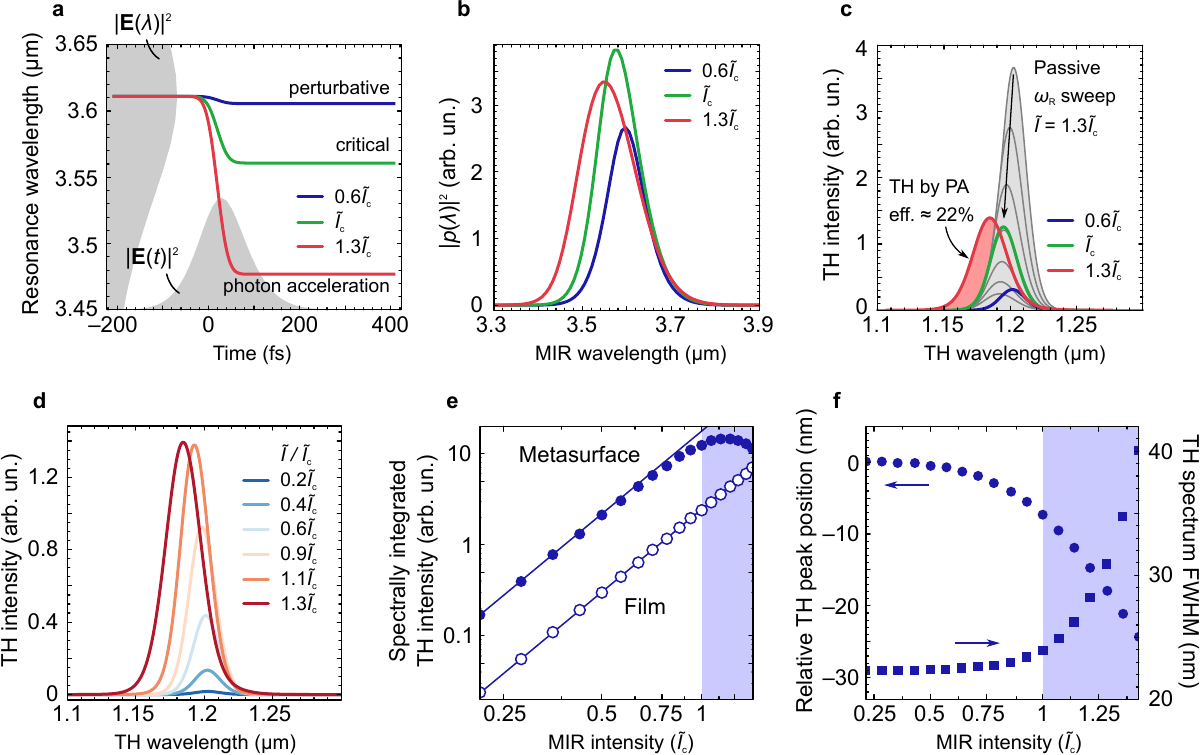}
\caption{\label{fig3} \textbf{Theoretical model of the PA-induced blue-shifted harmonics generation.}
\textbf{a}, Time-dependent resonant frequency of the metasurface due to 4PA FC generation plotted for three normalized peak intensities of the MIR pulse:
$\tilde I= 0.6 \tilde I_c$ (perturbative regime, blue curve),
$\tilde I= \tilde I_c$ (critical regime, green curve), and
$\tilde I = 1.3\tilde I_c$ (non-perturbative PA regime, red curve). Shaded areas: the frequency-domain (on the left) and time-domain (on the bottom) profiles of the incident pulse.
\textbf{b}, Spectra of the MIR electric field inside the metasurface. \textbf{c}, \textbf{d}, Predicted THG spectra for different input MIR intensities. In \textbf{c}, a comparison between the predictions of the models of time-dependent metasurface model (thick lines; coloring as in \textbf{a}) and of the time-independent (thin lines) metasurfaces. The family of time-independent metasurfaces satisfies $0< \Delta \omega_R/\omega_R^0 < 4\%$ and $\tilde I = 1.3\tilde I_c$. Filled red area: the portion of the spectrum inaccessible within the time-independent framework.
\textbf{e}, Spectrum-integrated THG intensity versus input intensity for the unpatterned film (open circles) and resonant metasurface (closed circles). THG from the film is simulated as from a significantly non-resonant structure with $\gamma_R=10^{15}$~s$^{-1}$ with an intensity calibration coefficient that is adjusted for good agreement with the experiment at low intensities.
Lines: guide-for-the-eye cubic dependences.
\textbf{f}, Intensity-dependent THG peak position (circles) and THG bandwidth (squares) generated by the metasurface. Shaded areas in \textbf{e} and \textbf{f}: the non-perturbative PA regime.}
\end{figure*}

\textbf{Theoretical model.}
The observed blue-shifting, broadening and saturation of the THG spectra can be explained by a simple coupled-mode theory (CMT) involving a single cavity mode with a mode amplitude
$p(t)$, characterized by its time-dependent resonant frequency $\omega_R(t) \equiv \omega_R^0 +\Delta\omega_R N(t)/N_{\rm max}$ and damping factor $\gamma_R(t) \equiv \gamma_R^0 +\Delta\gamma_R N(t)/N_{\rm max}$, and coupled to the incident optical field $\tilde E(t)$ according to~\cite{Haus1984,minkov2017}:
\begin{equation}
\frac{dp(t)}{dt} +[\imath \omega_R(t) +\gamma_R(t)]p(t)=\kappa \tilde E(t),
\label{eq:CMT}
\end{equation}
where $\kappa$ is the coupling constant. Here, we assume a Gaussian incident laser pulse with $\tilde E(t)=\sqrt{\tilde I}\exp\left(-\imath \omega_L t - t^2/\tau_L^2\right)$, where $\tau_L=105$~fs and $\omega_L = \omega_R^0$, and $\tilde I$ is the
intensity of the pump. The model does not aim to reproduce the peak intensity of the resulting harmonics, which is affected by the absolute normalization of the MIR pulse intensity. The resonant frequency/linewidth  shifts are of greater importance for quantitative understanding the PA's role, and their estimation is described below.

In Eq.(\ref{eq:CMT}), the unperturbed $\omega_R^0$ and $\gamma_R^0$ are obtained by fitting the transmission spectrum obtained with FTIR (see Fig.~\ref{fig1}b), the coupling constant $\kappa=\sqrt{2\gamma_R^0}$~\cite{Haus1984} is calculated by neglecting non-radiative losses in the absence of FCs, and $N_{\rm max}$ is set to be the maximum FC density achieved in the experiments. The carrier-induced shifts are assumed to be proportional to carrier density $N(t)<N_{\rm max}$ produced via the 4PA:
\begin{equation}\label{eq:carriers}
  N(t) = N_{\rm max} \frac{2\sqrt{2}}{\tau_L \tilde I^4_{\rm max} \sqrt{\pi}} \int\limits_{-\infty}^t |\tilde E(t^\prime)|^8dt^\prime,
\end{equation}
where $\tilde I_{\rm max}$ is the maximum intensity.
We further assume that  $\Delta\omega_R = 2\cdot10^{13}$~s$^{-1}$ and $\Delta\omega_R=2.5 \Delta\gamma_R$; these quantities
are close to those obtained by pump--probe measurements (see Supplementary Note~5). The resulting dynamic frequency sweep profiles are shown in Fig.~\ref{fig3}a for three characteristic intensities of
$\tilde I=0.6\tilde I_c$, $\tilde I_c$ and $1.3\tilde I_c$, corresponding to three different regimes of PASIM operation; the meaning of the characteristic critical intensity $\tilde I_c$ will be clarified below.

After numerically solving Eq.~(\ref{eq:CMT}) to obtain the enhanced near-fields inside the metasurface $\propto p(t)$, we calculate their spectra $|p(\lambda)|^2$ (see Fig.~\ref{fig3}b) and observe their significant blue-shifting and spectral broadening with the increasing incident intensity $\tilde{I}$. In the time domain, the latter feature translates into shorter and more intense bursts of the electric field in Si, thus giving rise to more intense nonlinear THG spectra $I_{\rm TH}(\lambda_{\rm TH})$ given by
\begin{equation}
I_{\rm TH}(\lambda_{\rm TH})\propto\left|\int\limits_{-\infty}^\infty p^{3}(t) e^{\imath \omega_{\rm TH} t}dt\right|^2,
\end{equation}
where $\lambda_{\rm TH}=2\pi c/\omega_{\rm TH}$.

The resulting upconversion spectra are displayed in Fig.~\ref{fig3}c,d for three values of the input light intensity $\tilde I$. In Fig.~\ref{fig3}c, we illustrate the three regimes of PASIM operation by comparing the spectra with those obtained for a set of fixed  $\omega_R$. At low intensities, e.g.,
$\tilde I=0.6\tilde I_c$ (blue curve), the shape of the upconverted spectrum does not differ from that produced by a time-independent metasurface with $\omega_R=\omega_R^0$. As the intensity increases, the upconverted light progressively blue-shifts, and for the critical intensity $\tilde I = \tilde I_c$  (green curve)  we start observing PA-induced photons that cannot be produced by any static metasurface with time-independent resonances. The transition to the PA regime is best illustrated for the intensity
$\tilde I = 1.3\tilde I_c$ (red curve in Fig.~\ref{fig3}c). It is instructive to attempt reproducing the nonlinear spectrum by using a set of stationary resonant frequencies/linewidths corresponding to different values of $N_j/N_{\rm max}$, where $1\leq j \leq 6$. We observe that, without invoking PA, it is impossible to reproduce the red curve in Fig.~\ref{fig3}c by any one of such static metasurfaces, or even by using a weighted average of their corresponding THG spectra. The reason for that is the existence of highly blue-shifted THG photons (see the red shaded area).

By comparing the theoretical spectra shown in Fig.~\ref{fig3}d with their experimental counterparts shown in Fig.~\ref{fig2}c, we find that the simple model of PA described by Eqs.(1--4) semi-quantitatively captures the key spectral features of the PASIM-based THG: simultaneous increase in intensity, blue-shift, and spectral width with increasing laser intensity. The model rather accurately captures the saturation of the THG as shown in Fig.~\ref{fig3}e, the value of the blue-shift ($\sim25$~nm), and the spectral broadening from the initial linewidth of the resonance ($\sim22$~nm) up to $\sim40$~nm, as shown in Fig.~\ref{fig3}f. Even though the dynamic resonance sweep is a very complex process, especially because of the highly localized nature of FC generation, we have demonstrated that our model can explain the main features of the experiment, and can be potentially used for optimizing future metasurface designs.

\begin{figure*}
\includegraphics[width=0.95\textwidth]{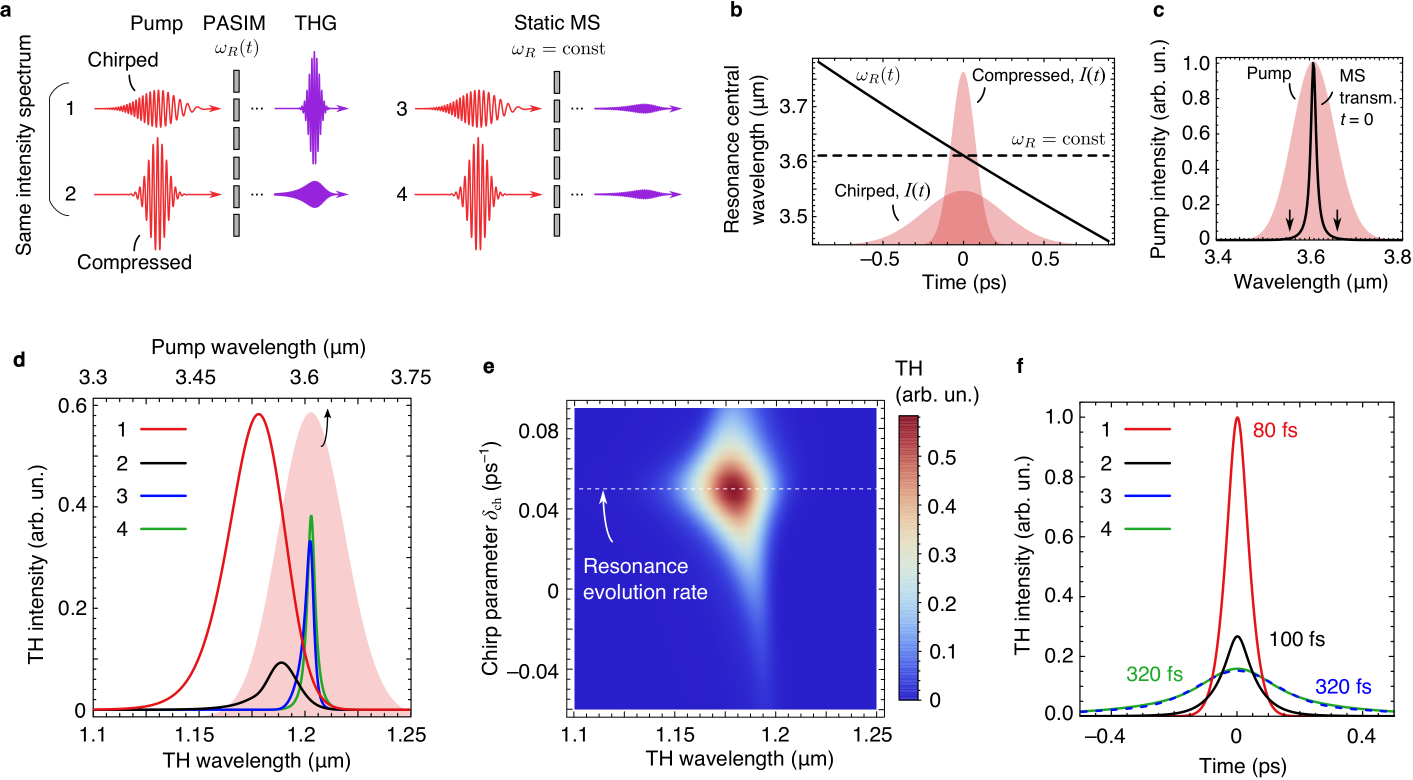}
\caption{
\label{fig4}
\textbf{Nonlinear interaction of broadband pulses with narrowband metasurfaces.}
\textbf{a}, THG from a narrow-band PASIM (cases 1 and 2) and static MS (cases 3 and 4) using broadband pulses, chirped or transform-limited (compressed).  
\textbf{b},
Evolution of the MS resonance frequency $\omega_R(t)$ for a PASIM (solid line) and a static MS (dashed line) superimposed with the temporal profiles (shaded areas) of the chirped ($\tau_L=500$~fs) and compressed ($\tau_L=150$~fs) MIR pulses with the same energy and spectral width.
\textbf{c},
Intensity spectrum of the pump (shaded area) and transmittance spectrum of the metasurfaces at $t=0$ (black curve): $\Delta \omega_L \approx 8 \Delta \omega_R$. The arrows indicate the sweep range of the PASIM's $\omega_R(t)$ within the temporal FWHM of the chirped MIR pulse.
\textbf{d},
Spectra of THG generated in cases 1--4. Shaded area referenced to the top axis: the spectrum of the MIR pulses.
\textbf{e}, THG emission from the PASIM for different values of the chirp parameter $\delta_{\rm ch}$. Both the bandwidth and efficiency peak at $\delta_{\rm ch} = \alpha$.
\textbf{f}, THG pulses generated in cases 1--4 after perfect compression, illustrating the possibility of generating ultrashort ($\tau_{\rm TH} < 80$~fs) bursts of the THG radiation by the PASIM. Metasurfaces' parameters: $\alpha=0.05$~ps$^{-1}$ (for PASIM), $\Delta \omega_R = 1$~ps$^{-1}$ (for both metasurfaces).
}
\end{figure*}

\section{Discussion}
We have estimated the energy portion of the upconverted radiation that is unreachable by a passive frequency sweep, $\eta_{\rm PA} \approx 22\%$, by calculating the shaded area  in Fig.~\ref{fig3}c relative to the area covered by the gray curves. 
Comparing to previous demonstrations of PA in microcavities \cite{Preble2007,Tanabe2009}, which yielded up to $\eta_{\rm PA}\approx50\%$, we note that the value of $\eta_{\rm PA}$ demonstrated here is achieved at much larger relative wavelength shifts (about $2.7\%$ of the central wavelength versus $<0.5\%$ shown before), and without any external pump. Moreover, the ultra-thin nature of the PASIM potentially enables tunable high-harmonics generation despite their finite absorption in Si. While comparable PA-associated blue shifts of the THG have been observed~\cite{Siders1996} in ionisable gases, the required laser intensities were in the PW/cm$^2$ range, i.e., almost $5$ orders of magnitude higher than those used in our experiments.

As an application example, we will discuss how the PA mechanism enables a new approach to improving photons' capture and nonlinear conversion: utilizing optical pulse shaping/chirping \cite{Weiner2000,Neyra2016} to engineer the right- and left-hand sides of Eq.~(\ref{eq:CMT}). We demonstrate that significant enhancement of the bandwidth and intensity of the THG can be achieved by matching the instantaneous frequency of a chirped laser pulse to the time-dependent resonant frequency of the metasurface. Figure \ref{fig4} shows the calculated THG spectra produced by the interaction of a time-dependent PASIM with $\omega_R(t)=\omega_R(0)(1+\alpha t)$ (where $\alpha=0.05$~ps$^{-1}$) with an incident chirped laser pulse whose electric field is given by
\begin{equation}
E_{\rm in}(t)= E_0 e^{- t^2/\tau^2_L} \times e^{-\imath \omega_L t} \times e^{-\imath \delta_{\rm ch} \omega_L t^2/2},
\end{equation}
where the last term describes a linear frequency chirp
with the normalized rate of change $\delta_{\rm ch}$. 
By choosing the FWHM of the pulses' intensity spectrum $\Delta \omega_L$, and the instantaneous width of the metasurfaces' resonance $\Delta \omega_R$ in such a way that $\Delta \omega_L \gg \Delta \omega_R$, we demonstrate that the benefits of the high-Q metasurface and a broadband incident laser pulse can be combined to achieve broadband THG with high conversion efficiency.
Four cases are considered in Fig.\ref{fig4}a: cases $1-2$ of a PASIM with a linearly evolving resonance frequency $\omega_R(t)=\omega_R(0)(1+\alpha t)$ excited by either a chirped or a transform-limited (compressed) MIR pulse, and cases $3-4$ of a static metasurface with $\omega_R(t)=\omega_R(0)$ excited by the same pulses. The corresponding parameters of the pulses and metasurface are listed in the caption of Fig.\ref{fig4}. The chirped and compressed pulses are chosen to have identical spectral intensities.

The predictions of the CMT calculations for these four cases are plotted in Fig.~\ref{fig4}d. The PASIM excited by both the chirped pulse (Case 1) and the compressed pulse (Case 2) produces broadband THG, with former being much more efficient than the latter. The nonlinear response of the PASIM strongly depends on $\delta_{\rm ch}$, including its sign. Specifically, the enhancement by the PASIM is the highest if the chirp parameter $\delta_{\rm ch}$ matches the evolution rate $\alpha$ of the PASIM's resonance. This can be clearly observed in Fig.\ref{fig4}e, where the THG spectra from the PASIM are plotted for different values of $\delta_{\rm ch}$. To our knowledge, the PASIM concept is the first proposal for a {\it chirp-sensitive} nonlinear metasurface. 
In contrast, the passive metasurface (Cases 3 and 4) produces narrow-band THG with low efficiency because they do not utilize the portion of the optical bandwidth that is outside of the metasurface bandwidth.

The importance of utilizing the entire bandwidth of the laser pulse can be most easily appreciated from the following observation. We predict that a compressed THG signal from a PASIM pumped by an optimally-chirped laser pulse can be as short as $\tau_{\rm TH} \approx 80$~fs despite the long lifetime of the mode $\tau_R \equiv 1/\gamma_R \approx 1$~ps. This is accomplished by using an additional passive dispersive element (e.g., a pair of compressing gratings) that turns a long chirped THG pulse emerging from the PASIM into a transform-limited short pulse while preserving its bandwidth. The comparison between the four cases discussed above is shown in Fig.\ref{fig4}f. It is apparent that only in Case 1 an intense short THG pulse is produced. These results establish that a high-$Q$ photon-accelerating metasurface can exhibit simultaneously efficient {\it and} broadband response, thus provides a path toward  surpassing the time-bandwidth limit found in systems with resonances \cite{Tsakmakidis2017}.

As an outlook, here we speculate on the possibilities of scaling our approach to other wavelength ranges, as well as outline several potential applications of PASIMs. Many applications benefit from efficient frequency conversion in the near-infrared and the visible. In order to scale the design down by a factor of 6, so as to push the PASIM operation down the visible, the smallest feature (the gap size) should be on the order of about 35 nm at the same height-to-gap aspect ratio of 2:1. Current e-beam resist technology can produce sub-20-nm resist feature sizes \cite{Grigorescu2009} that, in appropriate etching conditions, will result in the desired pattern. As far as the specific semiconductors, GaP is arguably the best candidate for applications of PA in the visible, for it has a large refractive index ($n = 3.2 - 4$) and band gap (2.25 eV) required for resonant operation using collective modes used in this work. The main obstacle to large frequency shifts at shorter wavelengths is the $\lambda^2$-scaling of the Drude term. However, this is not the only known term that affects the refractive index of semiconductors. In our previous work \cite{Shcherbakov2017}, GaAs was used near the band gap, where the refractive index was affected by band filling and band shrinkage effects \cite{Bennett1990}. Therefore, by judicious choice of materials and band gap engineering in ternary semiconductors, PA may be extended to near-IR and visible frequencies.

Below we briefly discuss a potential application of PASIMs: filling the spectral gap between high optical harmonics, with the objective of generating satellite-free isolated attosecond pulses---something that is currently accomplished using bulky optical components \cite{Louisy2015}. We propose to utilize the process of high-harmonic generation (HHG) by an MIR pulse which nonlinearly interacts and gets spectrally broadened by a time-dependent PASIM. The role of the PASIM is to sufficiently broaden the spectrum of the pulse so that adjacent ($N$'th and $N+1$'th, where $N \gg 1$) harmonics spectrally overlap. A similar PASIM based on a high-mobility semiconductor (GaAs) with $Q\sim 10^3$ would be utilized, and its resonance wavelength swept by $\Delta \lambda_R/\lambda_R \sim 0.1$, where $\lambda_L \approx \lambda_R = 3.6$~$\mu$m and $\tau_L^{comp} \sim 170$fs are the wavelength and FWHM duration of the transform-limited intense MIR pulse used for metasurface-enhanced HHG. Such sweep of the metasurface resonance frequency requires FC concentrations on the order of $N_{\rm FC}\sim 3\cdot10^{18}$~cm$^{-3}$ due to the low effective mass of electrons in GaAs \cite{Bennett1990}. In Supplementary Figure 8, we show that the spectrum of the $N=15$th harmonic generated from a PASIM is sufficiently broad to fill the spectral gap between the $N$th and $N+1$st harmonics. Because the PASIM-induced spectral broadening of the $N$th harmonic is proportional to $N$, and high harmonics up to $N = 32$ have already been produced in solids \cite{You2017}, we anticipate that PASIMs could play a pivotal role in obtaining resonant solid-state HHG continua for the generation of attosecond pulses in the extreme UV.

To summarize, we have provided the first demonstration of photon acceleration in ultrathin semiconductor metasurfaces by observing a blue-shifted third harmonic generation with central frequencies of up to 3.1$\omega_L$. Relative wavelength shifts as high as $2.7\%$ have been observed under moderate laser intensities owing to excitation of collective high-Q metasurface resonances. Using a coupled-mode theory with time-dependent mode parameters, we have validated our experimental findings and estimated the overall photon acceleration efficiency at around $\eta_{\rm PA} \approx 22\%$. In the measured spectra, new frequencies of up to $\approx3.4\omega_L$ have emerged, with the spectral intensity of up to eight orders of magnitude higher than the projected intensity from an unstructured silicon film.
These findings indicate that photon-accelerating nanostructures represent a novel time-dependent nonlinear photonic platform that can find various applications in novel pulsed light sources.

{\footnotesize

\section{Acknowledgements}

This work was supported by the Office of Naval Research (ONR) and by the Air Force Research Laboratory (AFRL). This work was performed in part at the Cornell NanoScale Facility, a member of the National Nanotechnology Coordinated Infrastructure (NNCI), which is supported by the National Science Foundation (Grant ECCS-1542081).

\section{Author contributions}

M.R.S. and G.S. conceived the idea. M.R.S. fabricated the sample and characterized it using FTIR spectroscopy. M.R.S., K.W., N.T. and E.C. performed the nonlinear optical measurements. Z.F. performed COMSOL simulations. M.R.S. performed CMT calculations. All authors contributed to the preparation of the manuscript.

\section{Competing  financial interests}

The authors declare no competing  financial interests.

\section{Methods}

\textbf{Sample fabrication and characterization}
Samples of silicon metasurfaces were fabricated at the Cornell Nanoscale Facility (CNF) from a silicon-on-insulator wafer (600~nm undoped Si device layer on top of a 460~$\mu$m top-grade sapphire from University Wafer) using the following recipe. The substrate was cleaned with acetone, isopropanol and O$_2$ plasma; PMMA 495 was spun to form a 400-nm-thick layer and baked for 15 min at 170$^\circ$C; PMMA 950k was spun to form a 100-nm-thick layer and baked for 15 min at 170$^\circ$C; E-spacer conducting layer was spun at 6000~rpm; the pattern was exposed at 1000~$\mu$C/cm$^2$ (JEOL 9500FS) and developed in MIBK:IPA 1:3 solution; a 60-nm-thick Cr mask was electron-beam-evaporated and lifted off in sonicated acetone for 1 minute; the pattern was transferred to the silicon layer through HBr reactive ion etch (Oxford Cobra). Finally, Cr was removed with the commercially available Cr wet etchant.

\textbf{Infrared spectroscopy}
Bruker Vertex 80 FTIR spectrometer was upgraded to an external transmittance spectroscopy setup as described elsewhere~\cite{Wu2014} that enables collimated MIR spectroscopy, with the MIR beam focused to a spot size of about 300~$\mu$m in diameter using a pinhole imaging technique. The transmitted beam was sent to the detector and Fourier-analyzed by the spectrometer. Normalization was done using the signal from the clear sapphire area.

\textbf{Nonlinear optical measurements} Setup schematic and a comprehensive description of the nonlinear optical measurements are given in Supplementary Figure~3 and Supplementary Note~3, respectively.

\textbf{Finite element simulations}
We used COMSOL Multiphysics to model the response of the metasurfaces (no free carriers) by defining the computational domain as a slab with the dimensions of $p_x\times p_y\times 3$~$\mu$m, where $p_x=2.1$~$\mu$m and $p_y=2$~$\mu$m. Periodic boundary conditions were used for the domain boundaries parallel to the $x-z$ and $y-z$ planes, and perfectly matched layers were used for the domain boundaries parallel to the $x-y$. The dimensions of the metasurface were chosen to match those obtained from the SEM image. Wavelength-independent refractive indices of $n_{\rm Si}=3.45$ and $n_{\rm Al_2O_3}=1.7$ were used for Si and sapphire, respectively.

}

\newpage 

\def\bibsection{\section*{\refname}}
\renewcommand\refname{Supplementary references}

\draft

\makeatletter
\renewcommand{\fnum@figure}{Supplementary Figure \thefigure}
\makeatother

\newcommand*\mycommand[1]{\texttt{\emph{#1}}}
\renewcommand{\thepage}{S\arabic{page}}
\renewcommand{\theequation}{S\arabic{equation}}
\renewcommand{\thefigure}{\arabic{figure}}

\setcounter{equation}{0}
\setcounter{page}{1}
\setcounter{figure}{0}

\textbf{Supplementary Note 1}

\textbf{High-Q Metasurface Design }

\begin{figure*}
\includegraphics[width=0.8\textwidth]{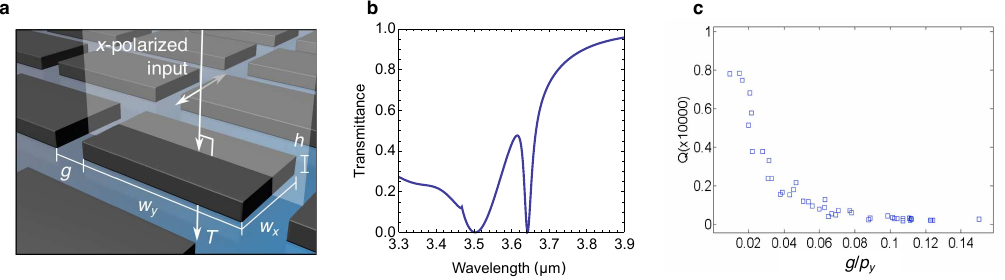}
\caption{\textbf{a}, The geometry of the PASIM with the dimensions and illumination parameters outlined. \textbf{b}, The calculated transmittance spectra of the PASIM used in the experiments. A sharp transmittance dip is seen at the coupling wavelength around $\lambda=3.64$~$\mu$m. \textbf{c}, $Q$-factor of the resonance dip as a function of the relative gap width $g/p_y$. For $p_y=2.1$~$\mu$m and $g=20$nm, the $Q$-factor can reach $Q\approx 8\times10^3$.}\label{hiQ}
\end{figure*}

The main considerations for the design of the photon accelerating semiconductor infrared metasurfaces (PASIMs) were outlined in similar realisations of high-Q semiconductor nanoparticle arrays [1,2]. The basic physics behind crafting a high-quality-factor metasurface is to create a planar waveguide system with a route to couple light in and out by periodic corrugations. A very simple  system we will consider here consists of conventional rectangular silicon waveguides with periodic through-notches that match the momentum of the initially normally incident light to a waveguide mode: $2\pi/p_y=\beta(\omega),$
where $p_y$ is the periodicity of corrugations and $\beta(\omega)$ is the propagation constant of the waveguide mode; see the structure of the metasurface in Supplementary Figure~\ref{hiQ}a. As a result, for the frequencies that satisfy the equation above, strong transmission dips are observed, as shown in Supplementary Figure~\ref{hiQ}b for a metasurface comprised of domino-shaped dielectric resonator antennas (DRAs). The geometry parameters of the metasurface, defined in the Supplementary Figure~\ref{hiQ}a, are as follows: $w_x=0.84$~$\mu$m, $w_y=1.82$~$\mu$m, $p_x=2$~$\mu$m, $p_y=2.1$~$\mu$m, and $h=600$~nm.
By changing the gap width $g$, one can tune the coupling between the DRAs, thereby affecting the $Q$-factor of the resonance. This effect is shown in Supplementary Figure~\ref{hiQ}c, where the $Q$ factor of the metasurface is plotted as a function of the gap size.

\textbf{Supplementary Note 2}

\textbf{Sample Fabrication}

The PASIM fabrication procedure is described in the Methods section of the main text of the paper. In Supplementary Figure~\ref{sample}(a), the fabrication flow process is illustrated. In Supplementary Figure~\ref{sample}(b), the choice of the sapphire substrate is justified by showing high transparency for mid-infrared radiation with $\lambda\approx3.6$~$\mu$m.

\begin{figure*}[b]
\includegraphics[width=0.8\textwidth]{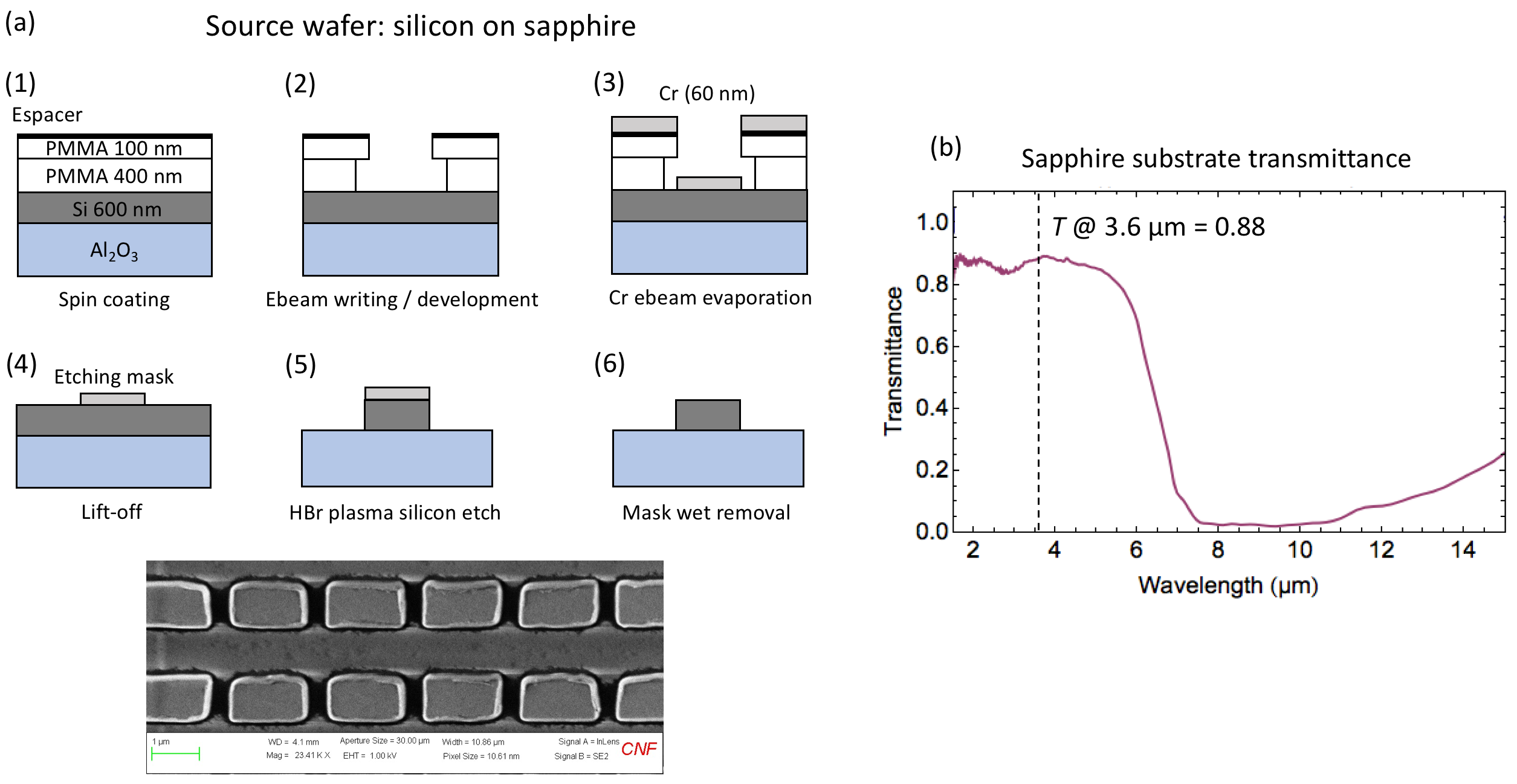}
\caption{(a) Sample fabrication process. (b) Transmittance of the sapphire substrate (purple curve) showing large transmittance in the spectral region around 3.6~$\mu$m.}\label{sample}
\end{figure*}

\textbf{Supplementary Note 3}

\textbf{Nonlinear-Optical Measurements}

In Supplementary Figure~\ref{setup_SI}, a schematic of the 	optical setup used for nonlinear measurements is shown.
The Extreme Mid-IR (EMIR) optical parametric amplifier (OPA) is a homebuilt KNbO$_3$/KTA 3-crystal/3-pass OPA. EMIR is pumped by The Ohio State University's GRAY laser, a homebuilt 80-fs Ti:Sapphire chirped pulse amplification system with a central wavelength of 780~nm and 4 mJ/pulse.
The repetition rate of EMIR can be varied nearly continuously between 1 and 500 Hz using an external Pockels-cell-based pulse picker.
EMIR was used to generate 200-fs mid-IR pulses with up to 40 $\mu$J/pulse. The output wavelength of EMIR can be varied continuously from $\lambda=2.7$ to 4.5~$\mu$m. For the experiments, the MIR (idler) beam was fixed at $\lambda=3.62$~$\mu$m. The 780~nm NIR, MIR, and $\lambda=1$~$\mu$m (signal) output beams are separated spatially, with the 1-$\mu$m signal being dumped and 780~nm pump being retained for use in pump--probe experiments. The residual NIR and MIR beams are roughly collimated to a size of about 2.5~mm. The NIR pulses were found to have a pulse duration of $200\pm15$~fs.

Output modes were characterized for several different wavelengths using a WinCamD-FIR2-16-HR 2 to 16~$\mu$m Beam Profiler System. Residual NIR pulse length was characterized using a BBO crystal based near-IR autocorrelator. MIR pulse duration was measured using an AGS-crystal-based MIR autocorrelator for 3 and 3.6~$\mu$m.

MIR spectra were obtained using a home-built spectrometer
based on a ThorLabs GR1325-30035 blazed ruled diffraction grating with a blaze wavelength of 3.5~$\mu$m as the dispersion element and the beam profiler sensor as the detector array. On the setup schematic, an inset demonstrates a typical image of the diffracted MIR beam. The spectrometer was calibrated with an A.P.E. Wavescan USB MIR spectrometer, which, due to a low sensitivity and operation speed, could not be used for the routine MIR spectroscopy.

For pump--probe and upconversion spectroscopy, the horizontally polarized MIR pulses first pass through a waveplate-polarizer assembly for precise energy control. The pulses travel through a variable delay line after which they are recombined with the NIR pulses via a dichroic mirror. The NIR pulses follow a separate but similar path. The collinear beams are focused using a CaF$_2$ $f=100$~mm plano-convex lens. In the sample plane, the spot sizes were found to be 300~$\mu$m FWHM for MIR and 400~$\mu$m FWHM for NIR. Both spots fit within the $500\times500$~$\mu$m structured area of the metasurfaces. The relative delay between MIR and NIR pulses was controlled dynamically using either the manual MIR delay line or the electronically controlled NIR delay line with sub-ps resolution. For self-tuning of the resonance, the NIR beam is blocked with a beam block. MIR fluences were varied from 1 to 6~mJ/cm$^2$ (see Table~1) and NIR fluences were varied from $<1$ to 4~mJ/cm$^2$ for the experiments. As a control method, a Si wafer was pumped in place of the sample. With NIR beam blocked, contamination of scattered light from the NIR and 1~$\mu$m signal was measured at the sample location. The NIR content was found to be 0.5~pJ per pulse and 1~$\mu$m signal was estimated to be of order 1~pJ/pulse. These pulse energies were determined to be insignificant to affect the sample during the experiment.

Upon transmission through the sample, the MIR beam and any upconversion signal
were collected with a CaF$_2$ $f = 50$~mm bi-convex lens. Any residual NIR was filtered using a Si window. In one configuration, a blazed grating/MIR camera combination is used as a high resolution MIR spectrometer. In another configuration, a commercial Ocean Optics NirQuest spectrometer (900--2500~nm) is used for detection of the upconverted radiation. THG signal was power-calibrated using the signal beam from the OPA at $\lambda=1.2$~$\mu$m that had a known power, after being attenuated by a set of neutral density filters with a known (measured) transmittance at this wavelength. By dividing the mean power of the THG beam by the mean power pump beam, an estimate maximum conversion efficiency of $10^{-9}$ was obtained.

\begin{table}[t]
\caption{Parameters of the mid-infrared beam: incident intensity $I_{\rm inc}$, incident fluence $F_{\rm inc}$, estimated equivalent intensity in the metasurface hotspot $I_{\rm hs}$.}
\begin{center}
\begin{tabular}{c|c|c|c}
& $I_{\rm inc}$ (GW/cm$^2)$ & $F_{\rm inc}$ (mJ/cm$^2$) & $I_{\rm hs}$ (TW/cm$^2$) \\
\hline Minimum & 5 & 1 & $< 1.75$ \\
\hline Maximum & 30 & 6 &  $< 10.5$
\end{tabular}
\end{center}
\label{default}
\end{table}%

\begin{figure*}
\includegraphics[width=0.75\textwidth]{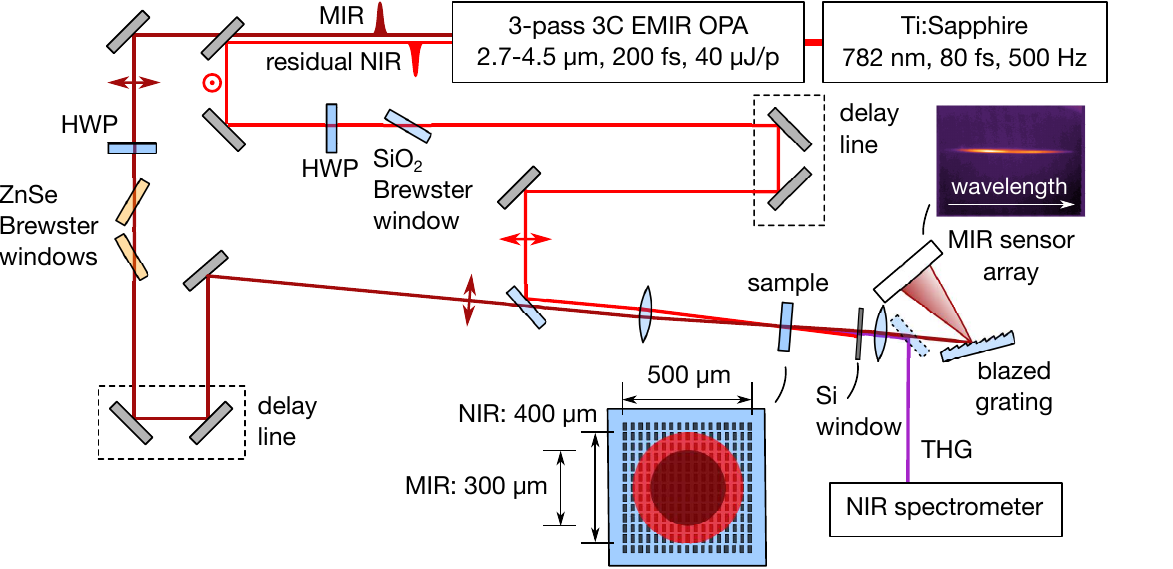}
\caption{Optical setup based on a Ti:Sapphire amplifier system coupled
to an optical parametric amplifier (OPA) capable of producing femtosecond laser pulses with the carrier wavelength tunable from 2.7 to 4.5~$\mu$m. Half-wave plates (HWP) are used for power management of both the near-IR pump beam, outlined with light red, and the mid-IR probe beam, outlined with dark red.}\label{setup_SI}
\end{figure*}

\textbf{Supplementary Note 4}

\textbf{Power-Dependent Transmittance of the PASIM Enabled by Four-Photon Absorption}

\begin{figure*}
\includegraphics[width=0.9\textwidth]{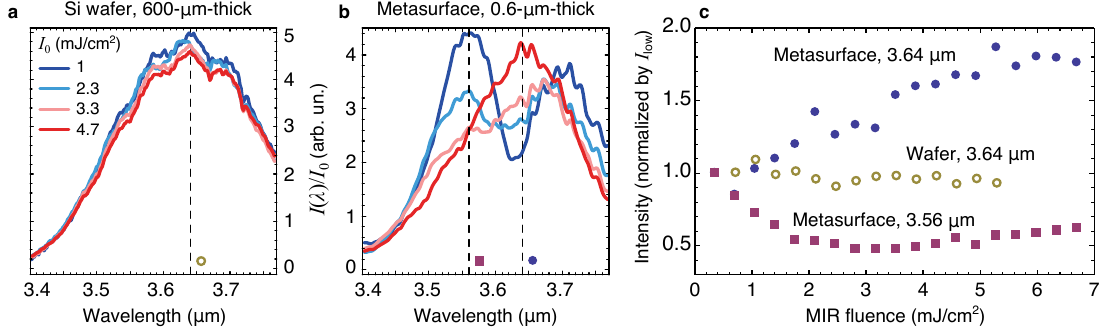}
\caption{\textbf{a}, Spectra of MIR pulses transmitted through a 500-$\mu$m-thick silicon wafer, divided by the input pulse fluence.
\textbf{b}, Spectra of MIR pulses transmitted through a 600-nm-thick silicon metasurface MS1, divided by the input pulse fluence. The same set of pump fluences is used as in panel \textbf{a}. A dip that starts off $\lambda=3.64$~$\mu$m corresponds to the high-Q resonance observed in the FTIR transmittance spectrum in Fig.~1b of the main text.
\textbf{c}, A comparison of the fluence-dependent transmittance on the resonance ($\lambda=3.64$~$\mu$m, closed circles) and off the resonance ($\lambda=3.56$~$\mu$m, closed squares) to that of the wafer at $\lambda=3.64$~$\mu$m (open circles).
}\label{fig2a}
\end{figure*}

The nonlinear properties of the PASIM can be interpreted only after the physics behind the resonance modification by mid-IR pulses is fully understood. This is  facilitated by measuring the transmitted pulse  spectra as a function of the incident pulse fluence (see Supplementary Figure~\ref{fig2a}) which controls the FC generation.  Note that undoped silicon has negligible linear absorbance [3] in the studied spectral range because the bandgap of silicon $E_g=1.12$~eV is more than three times larger than the MIR pulse carrier energy of $\hbar \omega = 0.35$~eV. Therefore, the interaction of MIR pulses with the metasurfaces is only affected by high-order processes, such as the four-photon absorption (4PA). We make estimates of the FC concentration via:
\begin{equation}
N = 2F(1-e^{-\alpha h})/4h E_{\rm pump},
\end{equation}
 where $F$ is the incident pulse fluence, $\alpha$ is the 4PA-induced absorption constant, $h$ is the thickness of the medium, and $E_{\rm pump}$ is the energy of a pump photon; the factor of 2 stands for a pair of generated FCs and the factor of 4 stands for the four photons needed for an absorption act. The 4PA-induced absorption is calculated with $\alpha = \beta_4 I^3$, where $\beta_4=3.5\cdot10^{-4}$~cm$^5$/GW$^{3}$ is the 4PA coefficient [4] and $I$ is the input intensity.
At an intensity of $I=10$~GW/cm$^2$ (fluence $F\approx2$~mJ/cm$^2$), in a bulk silicon wafer, $\alpha=3.5$~cm$^{-1}$, and the FC density is only $N \approx 3\cdot10^{16}$~cm$^{-3}$. Not surprisingly, essentially no self-modulation was observed for the $600$~$\mu$m-thick silicon wafer, as shown in Supplementary Figure~\ref{fig2}a for four fluence values.

However, the hot spots inside the metasurface (Fig.1a,b of the main text) enable a much higher FC density. With the local intensity of $I_{\rm local}=3.5$~TW/cm$^2$ (due to the $350$-fold intensity enhancement shown in Fig.~1b of the main text), the 4PA-induced absorption constant approaches $\alpha \approx 10^{10}$~cm$^{-1}$, which can induce full absorption of the incident pulse by the hot spots. Of course, Supplementary Eq.(1)  fails under high intensities due to pump depletion; however, these estimates suggest that giant self-modulations of MIR pulses are possible with our metasurfaces. Such fluence-dependent modifications of the resonance frequency and lifetime are shown in Supplementary Figure~\ref{fig2a}b. At low fluences (blue curve), the transmitted pulse contains a dip at $\lambda_{\rm dip}=3.64$~$\mu$m~$\approx \lambda_R$, which is in good agreement with the FTIR measurements. At higher fluences, the dip experiences considerable modifications, including blue-shift and broadening, and at the maximum fluence of $F=5.5$~mJ/cm$^2$, the dip is seen no more, as it has moved out of the spectral bandwidth of the mid-IR pulses.

Additionally, even at the lowest fluence, there is clear evidence of the PA. Specifically, the spectral component of the transmitted pulse are clearly redistributed from the dip region to a new {\it global peak} at $\lambda_{\rm peak}=3.56$~$\mu$m which is absent in the linear transmittance spectrum shown in Fig.~1b of the main text. This suggests that a significant fraction of the photons from the most populated portion of the original spectrum around $\lambda = \lambda_L$ are trapped by the metasurface resonance, and then accelerated to shorter wavelengths by the resonance blue-shifting during the FC generation. For this process to be efficient, femtosecond control of the FC generation is crucial. Specifically, it is important that significant FC generation takes place around the peak of the laser intensity. If it starts too early in the pulse, then by the arrival time $t_{\rm max}$ of the intensity maximum of the MIR pulse, the metasurface resonance frequency $\omega_R(t=t_{\rm max})$ will have shifted too far to the blue (where the number of the incident photons is small), and will have decreased its quality factor $Q_R(t=t_{\rm max})$ to the value that is insufficient for photon trapping. The temporal dynamics of the resonant frequency $\omega_R(t)$ is illustrated in Fig.3a of the main text.

\textbf{Supplementary Note 5}

\textbf{Plasma-Induced Blue-Shift and Damping of the Metasurface Resonance: Pump-Probe Experiments}

We can appreciate the values for the FC-induced metasurface resonance modification by measuring the MIR pulse transmittance spectra upon FC injection by an external NIR pump.
The spectra of the pulses transmitted through the sample were measured at different pump powers and/or pump--probe delays. Here we will concentrate on the cases of the pump preceding the probe by $\Delta \tau = 1$~ps to ensure that the probe interacts with a non-evolving metasurface whose resonance has been blue-shifted by the earlier pump pulse. Such choice of the pulse delays ensures that no photon acceleration takes place. To obtain the transmittance spectra, The acquired spectra were divided by the initial spectrum of the pulses transmitted through the substrate in the absence of the pump pulses . In order to extract the position of the resonance, we fitted the transmittance spectra near the dips by a parabolic dependence, which is the second-order Taylor expansion of the Lorentzian line shape. Several example of the curves after the division, along with the fit-extracted dip positions, is shown in Supplementary Figure~\ref{dip}. The maximum wavelength shift was observed to be $\Delta\lambda = 140$~nm (approximately a $4\%$ frequency shift), or $\Delta\omega_R=2\pi c\lambda^{-2}\Delta\lambda\approx1.8\cdot10^{13}$~s$^{-1}$.  The resonance damping factor was calculated from the manually defined FWHM of the resonance. Having taken into account the unperturbed (radiative) value of $\gamma_R^0=7\cdot10^{12}$~s$^{-1}$, the peak non-radiative increment to the resonance damping is estimated as $\Delta \gamma_R \approx 7.5\cdot10^{12}$~s$^{-1}$. In the modeling, to reasonably reproduce the experimental results, we found it necessary to use the following values:
$\Delta\omega_R=2\cdot10^{13}$~s$^{-1}$ and
$\Delta\gamma_R=5\cdot10^{12}$~s$^{-1}$,
which are fairly close to those obtained from the pump--probe measurements.

%

\begin{figure*}
\includegraphics[width=0.9\textwidth]{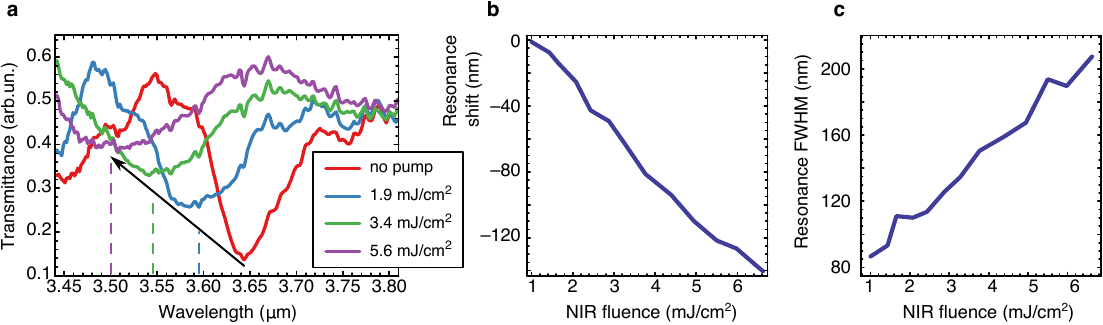}
\caption{
\textbf{a}, Pump-induced transmittance changes in the metasurfaces for several near-IR pump fluences color-coded according to the legend.
\textbf{b}, Retrieved resonance shift  as a function of the near-IR beam fluence.
\textbf{c}, Retrieved resonance FWHM as a function of the near-IR beam fluence.
}\label{dip}
\end{figure*}

\textbf{Supplementary Note 6}

\textbf{Blue-shifted Harmonics from an Off-resonant PASIM}

In addition to the experiments performed on the metasurface (referred to as the MS1) with an unperturbed resonant wavelength $\lambda_R^{(1)}$ coincident with the central wavelength of the MIR pulse $\lambda_L=3.62$~$\mu$m ($\lambda_R^{(1)} = \lambda_L$), we have also performed similar measurements on a second metasurface (referred to as the MS2). MS2 is designed to have a shorter unperturbed resonant wavelength $\lambda_R^{(2)}=3.56$~$\mu$m, i.e. even in the absence of FC generation the MS2 is blue-shifted with respect to $\lambda_L$. The measured linear transmittance through MS2 is shown in Supplementary Figure~6a. 
The power-dependent spectra of the non-integer THG for the MS2 are shown in the Supplementary Figure~6b. We observe that the peak of the THG spectrum is blue-shifted as the MIR fluence increases. Similar behavior was observed for the MS1 as well. There are, however, three key differences in the THG spectra collected from MS2 and MS1. 

First, we observe that the MS2 produces two spectral peaks (unlike the single peak for the MS1, see Fig.2c of the main manuscript). Because the long-wavelength peak is close to $\lambda_L/3$, we associate it with a non-resonant contribution to the THG from the detuned sample. Similarly to the unstructured film case, its position stays within a narrow $\approx 2$~nm range for all fluences. This peak is produced by the frequency trippling of the photons with the highest spectral density centered at $\lambda_L$. These photons do not undergo photon acceleration because they are not trapped by the metasurface. The main (short-wavelength) spectral peak is analogous to the blue-shifted peak observed in the experiments with the MS1. It is produced by the photons in the wings of the original MIR spectrum, whose wavelength coincides with $\lambda_R^{(2)}$ and is consequently shortened via photon acceleration. Therefore, the response of the MS2 combines the features of the resonant response of the MS1 (acceleration and frequency tripling of the trapped photons), and of the non-resonant response of the unpatterned Si film (non-resonant frequency tripling without photon acceleration).

\begin{figure*}
\includegraphics[width=0.9\textwidth]{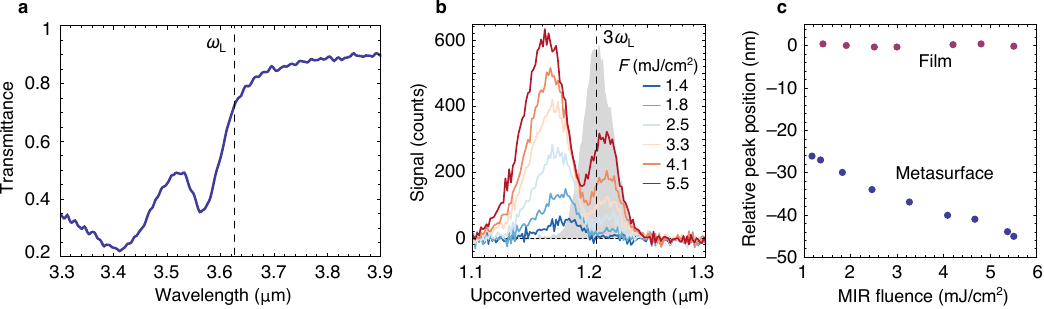}
\caption{Linear and nonlinear optical responses of the low-Q, off-resonant PASIM sample (MS2). \textbf{a}, Linear FTIR transmittance spectrum of the MS2. The dip at $\lambda_R^{(2)}=3.56$~$\mu$m indicates the excitation of a resonant collective mode of the metasurface. \textbf{b}, The spectra of the frequency-tripled signal as a function of the MIR fluence for the MS2. \textbf{c}, The position of the main upconversion peak in panel \textbf{b} as a function of the MIR fluence (blue dots). The same dependence measured for an unstructured Si film is shown with purple dots.
}\label{MS2}
\end{figure*}

Second, the peak spectral intensity of the THG from the MS2 corresponding to the highest MIR fluence is smaller by a factor $2$ that the corresponding peak intensity from the MS1. This reduction of the nonlinear signal is due to two factors: (i) the lower $Q$-factor translates into lower field enhancement, and (ii) there are fewer photons that are captured and accelerated by the evolving metasurface because of the initial blue shift of the MS2's resonance with respect to the spectral density peak of the incident MIR pulse.  

Finally, the blue-shifting of the central position of the THG peak does not appear to saturate even at the highest MIR fluence. as a function of fluence. This indicates that fewer FCs are produced inside the hot spots of the MS2 at a given fluence when compared with the FCs' density in the MS1. Therefore, the distortion of the spatial mode's profile remains small for the MS2, even at the highest MIR fluence.

\textbf{Supplementary Note 7}

\textbf{On Kerr and Thermo-optic Effects}

In a semiconductor subjected to an intense laser pulse, the main contributions to the refractive index modulation are photogenerated free carriers, Kerr effect and thermo-optic effect. In this work, we assume that the main contribution to the refractive index is the one from the photogenerated free carriers.
There are two main reasons why Kerr and thermo-optic effects can be assumed smaller than that from free carriers. 

First, in silicon, both Kerr and thermo-optic additions to the refractive index are positive: $\Delta n_{\rm Kerr}=n_2 I,$ where $n_2\approx3\times10^{-14}$~cm$^2$/W [4], $\Delta n_{\rm therm}=T dn/dT$, where $dn/dT=2\times10^{-4}$~1/K [5].
Positive $n_2$ and $dn/dT$ mean the anticipated resonance shifts are to the red part of the spectrum; instead, only blue shifts are observed. 

Second, thermal effects are known to affect the optical properties of materials after the system has thermalized, which usually happens several picoseconds after the pulse arrives. Since we are using a low-repetition-rate source, the system cools down before the next pulse comes, ensuring that thermo-optic effects do not affect our measurements. 

Interestingly though, the Kerr effect can be important at the mode hotspots, as 
 estimated from the known values of $n_2$ for Si. We arrive at $\Delta n_{\rm Kerr}\approx0.1$ at an input intensity of 10 GW/cm$^2$, if the local-field-induced equivalent intensity in the hot spot of the metasurface is 350 times that of the input field. This is on the same order with the estimated $\Delta n_{\rm FC}\approx -0.1$ we experimentally observe at this intensity. However, due to the observed blue-shifts, and potential FC-induced damping of the hotspot intensity, we believe the FC contribution dominates the response of the metasurface. Nevertheless, we predict that Kerr nonlinearities may start playing an important role under different, yet realistic experimental conditions, such as longer wavelengths  and lower intensities.

\textbf{Supplementary Note 8}

\textbf{Derivation of Eq.(3)}

Eq.(3) of the main text is a result of a well-established rate equation for four-photo-absorption-induced FC generation (see, for instance, Eq.(1) from [6]):
\begin{equation}
\frac{dN}{dt}=KI^4(t),
\end{equation}
where $K$ is a coefficient proportional to the 4PA constant. Then, we integrate it  assuming the input pulse is Gaussian:
\begin{equation}
N(t)=K\int\limits_{-\infty}^tI^4(t^\prime)dt^\prime,
\end{equation}
where $I=|E_{\rm in}(t)|^2=\tilde I \exp(-2t^2/\tau^2_L)$. Finally, to find $K$, we use:
\begin{equation}
N(+\infty)=N_{\rm max}\left(\frac{\tilde I}{\tilde I_{\rm max}}\right)^4.
\end{equation}

This equation ensures (a) that the FC density after the pulse is gone is proportional to $\tilde I^4$, and (b) that at $\tilde I= \tilde I_{\rm max}$, $N(+\infty)=N_{\rm max}$. From:
\begin{equation}
N(t)=K\int\limits_{-\infty}^{\infty}I^4(t^\prime)dt^\prime=K\tilde I^4 \sqrt{\frac{\pi \tau_L}{8}},
\end{equation}

applying Eqs.(S4,S5) to Eq.(S3), one comes up with Eq.(3) of the main text. 

\textbf{Supplementary Note 9}

\textbf{Reproduction of Fig.1d by CMT}

\begin{figure}[t!]
\includegraphics[width=\columnwidth]{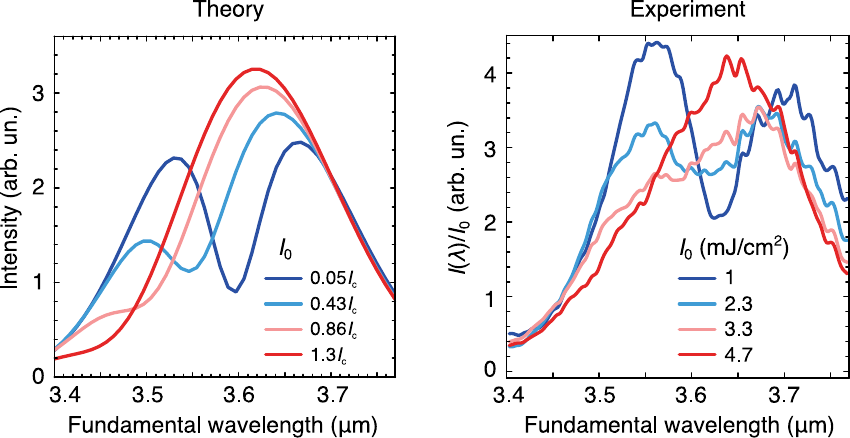}
\caption{Left: CMT results for pulse transmitted light spectra as a function of incident power. The parameters of the model are the same as those in Fig.3 of the main text, except for here, the transmitted fields are given instead of THG, as well as the out-coupling constant is let to be a free parameter. Right: corresponding experimental results, Fig.1d of the main text.
}\label{1d}
\end{figure}

Here, we used the same set of parameters we used to obtain Fig.3 of the main text, with the exception that the transmitted fundamental pulse spectrum is calculated instead of the THG. Output fields are calculated via $E_{\rm out}(t)=E_{\rm in}-\kappa p(t)$ where the coupling constant $\kappa=\sqrt{2\gamma}=3.7$~ps$^{-1/2}$, then the quantity $E_{\rm out}(t)$ is Fourier-transformed to $E_{\rm out}(\omega)$, and $|E_{\rm out}(\omega)|^2$ plotted for four different intensities in Suppl. Fig.~7. We find similar behavior in our theory as in the measured curves: for the lowest intensity, there is a dip in the center of the initially-Gaussian pulse spectrum; as intensity increases, the dip blue-shifts and broadens; at the maximum intensity, the dip is broadened and blue-shifted considerably away from its initial position. 

\textbf{Supplementary Note 10}

\textbf{Broadband High Harmonic Generation}

The ability of PASIMs to manipulate the bandwidth of the pulsed radiation and its optical harmonics can be applied for generation of high harmonic continuum and for spectral manipulation. 

Specifically, here, we theorize an experimentally realistic setting where PASIMs show broadband generation of high optical harmonics enough to spectrally overlap the neighboring orders and produce broadband radiation in the UV. We consider a metasurface based on a high-mobility semiconductor (GaAs) with a resonance Q-factor of $10^3$. The central wavelength of the resonance is being swept by 0.1 within the fwhm duration of the pulse around a value of 3.2~$\mu$m, while the (chirped) pulse duration is assumed to be 170~fs. Such a sweep is predicted to be possible at maximum FC plasma concentrations of around $N_{\rm FC}\approx 3\cdot10^{18}$~cm$^{-3}$ [7]. Note that at such FC densities, FC absorption can start playing detrimental role on the quality factor of the resonance. Although at the chosen wavelength, the absorption constant is on the order of $<100$~cm$^{-1}$ ($n^{\prime\prime} < 0.003$), which we think is insufficient to significantly impede the operation of the cavity, further studies are needed to pinpoint to role of FC absorption and its mitigation.

In Supplementary Figure 8 we show the spectra of the 15th and 16th harmonics emitted from the time-varying metasurface pumped by the chirped pulse: the photon-accelerated emission spectrum consists of the harmonic continuum. Since the amount of blue-shift in the $N$th harmonic grows as a power of $N$, and HHG of up to $N = 32$ has been observed in solids [8], we find this result to be an important step toward HHG UV continuum generation, and one can expect metasurface cavities to play a pivotal role in attosecond metrology in the extreme UV.

\begin{figure}[t!]
\includegraphics[width=0.3\textwidth]{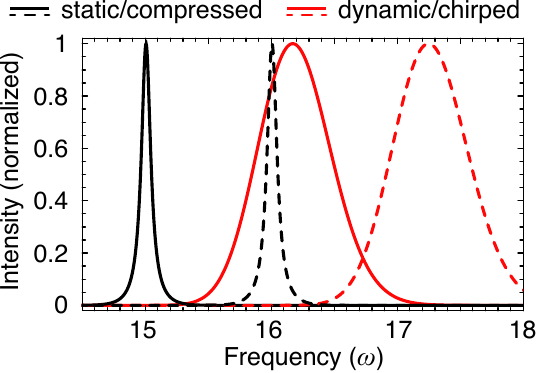}
\caption{
Spectra of the 15th (solid lines) and 16th (dashed lines) optical harmonics generated by a static metasurface and a compressed femtosecond pulse (black) and a PASIM and a properly chirped pulse (red), revealing the spectral overlap of the accelerated harmonics. 
}\label{1d}
\end{figure}

\textbf{Supplementary Note 11}

\textbf{Role of $\gamma \tau$ parameter and high Q-factors in efficient PA}

We start with a single-mode CMT:
\begin{equation}
\frac{da(t)}{t}+[i\omega(t) +\gamma]a(t)=\sqrt{\gamma} s(t),
\end{equation}
where $a(t)$ is the mode amplitude, $\omega(t)$ is its eigenfrequency that we assume time-dependent, $\gamma$ is its damping constant that we assume fixed, and $s(t)$ is the excitation field. We assume our system has only one mode, which is a good approximation for multimode systems that evolve slowly so as to be orthogonal at any given time. The exact solution of the CMT equation is:
\begin{equation}
a(t)=\sqrt{\gamma}\int\limits_{-\infty}^t dt^\prime e^{-\gamma(t-t^\prime)-i[\phi(t)-\phi(t^\prime)]}s(t^\prime),
\end{equation}
where $\phi(t)=\int_{-\infty}^t \omega(\xi)d\xi$ is the phase advance of the mode, and the initial condition $a(-\infty)=0$ was used. Typically, the excitation will come in a form of a pulse:
\begin{equation}
s(t)=A(t)e^{-i\psi(t)}.
\end{equation}
Here, $A(t)$ is a real, slowly varying envelope function that satisfies $\lim\limits_{t\rightarrow \pm\infty}{A(t)}=0$, and $\psi(t)$ is the phase of the excitation. For a given $\omega(t)$ there exists an optimal evolution of the excitation phase:
\begin{equation}
\psi(t)=\int\limits_{-\infty}^t \omega(\xi)d\xi+i\varphi
\end{equation}
Under the optimal excitation phase, the solution takes the form of:
\begin{equation}
a(t)=\sqrt{\gamma}e^{-\gamma t-i\phi(t)}
\int\limits_{-\infty}^t dt^\prime e^{\gamma t^\prime}A(t^\prime),
\end{equation}
Let us assume that the output radiation is defined as $s_+(t)=\sqrt{\gamma}a(t)$. If the lifetime of the cavity is small with respect to the pulse duration $\tau$, or $\gamma\tau\gg1$, from Eq.(S10), we can assume $e^{\gamma t^\prime}$ to be a slowly varying function with respect to  $A(t)$, so that:
\begin{equation}
s_+(t)\approx A(t)e^{-i\psi(t)}=s(t),
\end{equation}
and no frequency conversion takes place. It means that the most interesting regime for frequency synthesis is being $\gamma\tau$ on the order of or smaller than 1, meaning high-Q resonators produce more PA photons than the low-Q ones.

To illustrate the importance of a high-Q metasurface for producing measurable photon acceleration, we have carried out numerical simulations of the Eq.(2) of our manuscript for two cases of time-varying resonators: (i) a high-Q resonators ($Q=70$), and (ii) a low-Q resonator ($Q=7$). In both cases, a Gaussian transform-limited laser with the central wavelength of $\lambda_L=3.61$~$\mu$m and duration $\tau_L=200$~fs was used as an input. These two cases, respectively, correspond to the resonator bandwidth being within the laser bandwidth (the PA case described in our manuscript), and the opposite case of the laser bandwidth being entirely within the resonator bandwidth (the case described in Refs.[52, 53] of the main text). Two time protocols for varying the resonant wavelengths of the metasurface were chosen: (a) static metasurface, and (b) a metasurface with a resonant wavelength swept from $\lambda_R(-\infty)=3.61$~$\mu$m to $\lambda_R(+\infty)=3.44$~$\mu$m. Therefore, both the broadband and narrow-band time-varying metasurfaces are swept over an identical wavelength range that exceeds the linewidth of the incident MIR laser pulse.

\begin{figure*}[t!]
\includegraphics[width=0.8\textwidth]{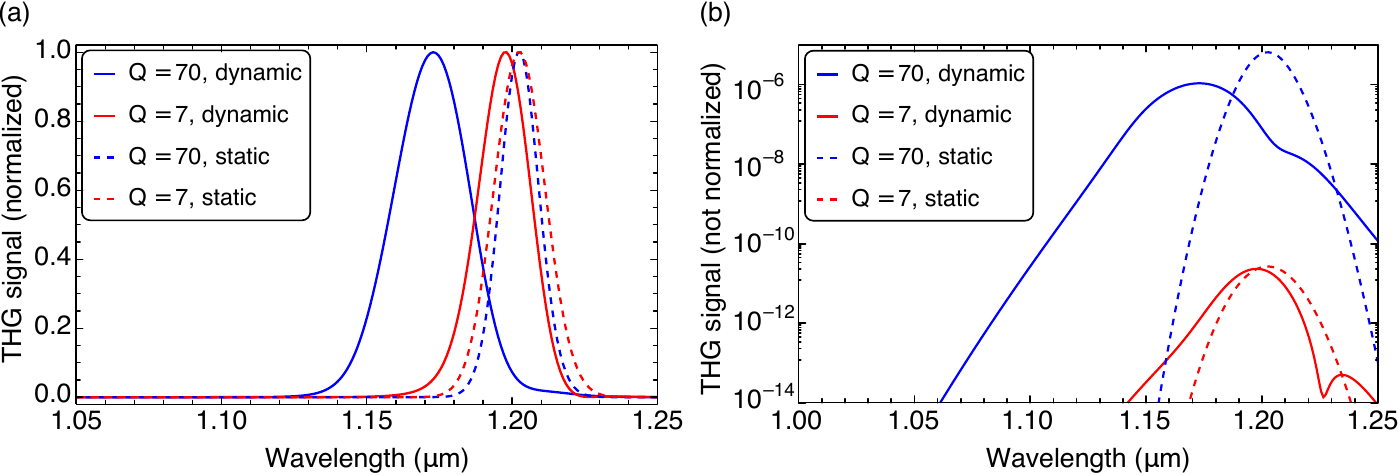}
\caption{
THG interaction of an input MIR pulse with high- and low-Q resonators that are either time-dependent (dynamic) or time-independent (static). MIR pulse parameters: $\lambda_L=3.61$~$\mu$m, $\tau_L=200$~fs. The resonant frequencies of the dynamic resonators are swept from $\lambda_R(-\infty)=3.61$~$\mu$m to $\lambda_R(+\infty)=3.44$~$\mu$m. The largest spectral broadening and blue-shifting of the THG pulse is obtained using a high-Q dynamic resonator (solid blue line). Clearly, the blue-shift and the spectral broadening achieved using a high-Q metasurface are much more prominent than those in the case of a low-Q metasurface. (a) Normalized THG signal in the linear scale, (b) THG signal before normalization in the logarithmic scale.
}\label{1d}
\end{figure*}

In Supplementary Figure 9, the normalized spectra of the near-IR pulses produced via THG mechanism are plotted for all four cases. The low-Q metasurface, which is used here to emulate the broad-band resonator of Refs.[52,53] of the main text, produces a spectrally-reshaped laser pulse (red solid line) that appears to be slightly on the blue side of the THG pulse produced from a static resonator (red dashed line). The spectral shift of order  is almost an order of magnitude smaller than the spectral shift of the resonant frequency. The strong spectral overlap between the THG spectra from the static (red dashed line) and dynamic (red solid line) spectra is a clear proof that photon acceleration can be neglected in the case of a broadband (low-Q) resonator.  In stark contrast, the high-Q resonator, which is used here to emulate the narrow-band metasurfaces used in our experiments, produces a broadband and strongly blue-shifted optical pulse (solid blue line). A significant fraction of frequency-tripled photons exist in the part of the spectrum (e.g., $\lambda_{THG}<1.15$~$\mu$m) where no progenitor photons ($\lambda_{THG}<3.45$~$\mu$m) exist in the original input pulse. Therefore, no amount of spectral filtering by a resonator with a blue-shifted resonance frequency could have given rise to the THG photons. Additionally, we find that narrow-band (high-Q) resonators produce nonlinear responses that are several orders of magnitude stronger than their broadband (low-Q) counterparts. Unlike the normalized spectra plotted Supplementary Figure 9(a), the absolute magnitudes of the THG signals are plotted in Supplementary Figure 9(b) for the two resonators. The advantage of using temporally changing high-Q resonators is obvious: the resulting THG signals are not only spectrally broader than their low-Q counterparts, but also 6 orders of magnitude stronger.

\vspace{0.3cm}

\textbf{SUPPLEMENTARY REFERENCES}

[1] Yang et al., Nat. Commun. \textbf{5}, 5753 (2014).

[2] Parry et al., Appl. Phys. Lett. \textbf{111}, 053102 (2017).

[3] Soref et al., J. Opt. A \textbf{8}, 840 (2006).

[4] Gai et al., Laser Photon. Rev. \textbf{7}, 1054 (2013).

[5] Komma et al., Appl. Phys. Lett. \textbf{101}, 041905 (2012).

[6] Hasselbeck et al., J. Opt. Soc. Am. B \textbf{14}, 1616 (1997). 

[7] Osamura and Murakami, Jap. J. Appl. Phys. \textbf{11}, 365 (1972).

[8] You et al., Nat. Commun. \textbf{8}, 724 (2017).

[9] Haus, \textit{Waves And Fields In Optoelectronics}(Prentice-Hall, 1984).

[10]	Fan et al., J. Opt. Soc. Am. A \textbf{20}, 569 (2003).

%
%
%


\begin{thebibliography}{10}
\expandafter\ifx\csname url\endcsname\relax
  \def\url#1{\texttt{#1}}\fi
\expandafter\ifx\csname urlprefix\endcsname\relax\def\urlprefix{URL }\fi
\providecommand{\bibinfo}[2]{#2}
\providecommand{\eprint}[2][]{\url{#2}}

\bibitem{Franken1961}
\bibinfo{author}{Franken, P.~A.}, \bibinfo{author}{Hill, A.~E.},
  \bibinfo{author}{Peters, C.~W.} \& \bibinfo{author}{Weinreich, G.}
\newblock \bibinfo{title}{{Generation of optical harmonics}}.
\newblock \emph{\bibinfo{journal}{Phys. Rev. Lett.}}
  \textbf{\bibinfo{volume}{7}}, \bibinfo{pages}{118--119}
  (\bibinfo{year}{1961}).

\bibitem{Maker1986}
\bibinfo{author}{Maker, P.} \& \bibinfo{author}{Terhune, R.}
\newblock \bibinfo{title}{{Study of Optical Effects Due to an Induced
  Polarization Third Order in the Electric Field Strength}}.
\newblock \emph{\bibinfo{journal}{Phys. Rev.}} \textbf{\bibinfo{volume}{137}},
  \bibinfo{pages}{344--350} (\bibinfo{year}{1965}).

\bibitem{Burnett1977}
\bibinfo{author}{Burnett, N.~H.}, \bibinfo{author}{Baldis, H.~A.},
  \bibinfo{author}{Richardson, M.~C.} \& \bibinfo{author}{Enright, G.~D.}
\newblock \bibinfo{title}{{Harmonic generation in CO2 laser target
  interaction}}.
\newblock \emph{\bibinfo{journal}{Appl. Phys. Lett.}}
  \textbf{\bibinfo{volume}{31}}, \bibinfo{pages}{172--174}
  (\bibinfo{year}{1977}).

\bibitem{Yamada1993}
\bibinfo{author}{Yamada, M.}, \bibinfo{author}{Nada, N.},
  \bibinfo{author}{Saitoh, M.} \& \bibinfo{author}{Watanabe, K.}
\newblock \bibinfo{title}{{First-order quasi-phase matched LiNbO3 waveguide
  periodically poled by applying an external field for efficient blue
  second-harmonic generation}}.
\newblock \emph{\bibinfo{journal}{Appl. Phys. Lett.}}
  \textbf{\bibinfo{volume}{62}}, \bibinfo{pages}{435--436}
  (\bibinfo{year}{1993}).

\bibitem{Campagnola2003}
\bibinfo{author}{Campagnola, P.~J.} \& \bibinfo{author}{Loew, L.~M.}
\newblock \bibinfo{title}{{Second-harmonic imaging microscopy for visualizing
  biomolecular arrays in cells, tissues and organisms}}.
\newblock \emph{\bibinfo{journal}{Nature Biotechnol.}}
  \textbf{\bibinfo{volume}{21}}, \bibinfo{pages}{1356--1360}
  (\bibinfo{year}{2003}).

\bibitem{Debarre2006}
\bibinfo{author}{D{\'{e}}barre, D.} \emph{et~al.}
\newblock \bibinfo{title}{{Imaging lipid bodies in cells and tissues using
  third-harmonic generation microscopy}}.
\newblock \emph{\bibinfo{journal}{Nature Methods}}
  \textbf{\bibinfo{volume}{3}}, \bibinfo{pages}{47--53} (\bibinfo{year}{2006}).

\bibitem{Heinz1982}
\bibinfo{author}{Heinz, T.~F.}, \bibinfo{author}{Chen, C.~K.},
  \bibinfo{author}{Ricard, D.} \& \bibinfo{author}{Shen, Y.}
\newblock \bibinfo{title}{{Spectroscopy of Molecular Monolayers by Resonant
  Second-Harmonic Generation}}.
\newblock \emph{\bibinfo{journal}{Phys. Rev. Lett.}}
  \textbf{\bibinfo{volume}{48}}, \bibinfo{pages}{478--481}
  (\bibinfo{year}{1982}).

\bibitem{Shen1989}
\bibinfo{author}{Shen, Y.~R.}
\newblock \bibinfo{title}{{Surface properties probed by second-harmonic and
  sum-frequency generation}}.
\newblock \emph{\bibinfo{journal}{Nature}} \textbf{\bibinfo{volume}{337}},
  \bibinfo{pages}{519--525} (\bibinfo{year}{1989}).

\bibitem{Shurik2011}
\bibinfo{author}{Stiopkin, I.} \emph{et~al.}
\newblock \bibinfo{title}{{Hydrogen bonding at the water surface revealed by
  isotopic dilution spectroscopy}}.
\newblock \emph{\bibinfo{journal}{Nature}} \textbf{\bibinfo{volume}{474}},
  \bibinfo{pages}{192--195} (\bibinfo{year}{2011}).

\bibitem{Mendonca2000}
\bibinfo{author}{Mendon{\c{c}}a, J.~T.}
\newblock \emph{\bibinfo{title}{{Theory of Photon Acceleration}}}
  (\bibinfo{publisher}{Insitute of Physics Publishing},
  \bibinfo{address}{Bristol and Philadelphia}, \bibinfo{year}{2000}).

\bibitem{Yablonovitch1974}
\bibinfo{author}{Yablonovitch, E.}
\newblock \bibinfo{title}{{Self-Phase Modulation of Light in a Laser-Breakdown
  Plasma}}.
\newblock \emph{\bibinfo{journal}{Phys. Rev. Lett.}}
  \textbf{\bibinfo{volume}{32}}, \bibinfo{pages}{1101--1104}
  (\bibinfo{year}{1974}).

\bibitem{Wilks1989}
\bibinfo{author}{Wilks, S.~C.}, \bibinfo{author}{Dawson, J.~M.},
  \bibinfo{author}{Mori, W.~B.}, \bibinfo{author}{Katsouleas, T.} \&
  \bibinfo{author}{Jones, M.~E.}
\newblock \bibinfo{title}{{Photon accelerator}}.
\newblock \emph{\bibinfo{journal}{Phys. Rev. Lett.}}
  \textbf{\bibinfo{volume}{62}}, \bibinfo{pages}{2600--2603}
  (\bibinfo{year}{1989}).

\bibitem{Felsen1970}
\bibinfo{author}{Felsen, L.~B.} \& \bibinfo{author}{Whitman, G.~M.}
\newblock \bibinfo{title}{{Wave Propagation in Time-Varying Media}}.
\newblock \emph{\bibinfo{journal}{IEEE Trans. Antennas Propag.}}
  \textbf{\bibinfo{volume}{AP-18}}, \bibinfo{pages}{242--253}
  (\bibinfo{year}{1970}).

\bibitem{Wood1991}
\bibinfo{author}{Wood, W.}, \bibinfo{author}{Siders, C.} \&
  \bibinfo{author}{Downer, M.}
\newblock \bibinfo{title}{{Measurement of Femtosecond Ionization Dynamics of
  Atmospheric Density Gases by Spectral Blueshifting}}.
\newblock \emph{\bibinfo{journal}{Phys. Rev. Lett.}}
  \textbf{\bibinfo{volume}{67}}, \bibinfo{pages}{3523--3536}
  (\bibinfo{year}{1991}).

\bibitem{Savage1992}
\bibinfo{author}{Savage, R.~L.}, \bibinfo{author}{Joshi, C.} \&
  \bibinfo{author}{Mori, W.~B.}
\newblock \bibinfo{title}{{Frequency upconversion of electromagnetic radiation
  upon transmission into an ionization front}}.
\newblock \emph{\bibinfo{journal}{Phys. Rev. Lett.}}
  \textbf{\bibinfo{volume}{68}}, \bibinfo{pages}{946--949}
  (\bibinfo{year}{1992}).

\bibitem{Yablonovitch1973}
\bibinfo{author}{Yablonovitch, E.}
\newblock \bibinfo{title}{{Spectral broadening in the light transmitted through
  a rapidly growing plasma}}.
\newblock \emph{\bibinfo{journal}{Phys. Rev. Lett.}}
  \textbf{\bibinfo{volume}{31}}, \bibinfo{pages}{877--879}
  (\bibinfo{year}{1973}).

\bibitem{Siders1996}
\bibinfo{author}{Siders, C.~W.} \emph{et~al.}
\newblock \bibinfo{title}{{Blue-shifted third-harmonic generation and
  correlated self-guiding during ultrafast barrier suppression ionization of
  subatmospheric density noble gases}}.
\newblock \emph{\bibinfo{journal}{J. Opt. Soc. Am. B}}
  \textbf{\bibinfo{volume}{13}}, \bibinfo{pages}{330--335}
  (\bibinfo{year}{1996}).


  \bibitem{Turchinovich2012}
\bibinfo{author}{Turchinovich, D.} \emph{et~al.}
\newblock \bibinfo{title}{{Self-phase modulation of a single-cycle terahertz pulse by nonlinear free-carrier response in a
semiconductor}}.
\newblock \emph{\bibinfo{journal}{Phys. Rev. B}}
  \textbf{\bibinfo{volume}{85}}, \bibinfo{pages}{201304(R)}
  (\bibinfo{year}{2002}).


  \bibitem{Blanco2014}
\bibinfo{author}{Blanco-Redondo, A.} \emph{et~al.}
\newblock \bibinfo{title}{{Observation of soliton compression in silicon photonic crystals}}.
\newblock \emph{\bibinfo{journal}{Nat. Commun.}}
  \textbf{\bibinfo{volume}{5}}, \bibinfo{pages}{3160}
  (\bibinfo{year}{2014}).


\bibitem{Preble2007}
\bibinfo{author}{Preble, S.~F.}, \bibinfo{author}{Xu, Q.} \&
  \bibinfo{author}{Lipson, M.}
\newblock \bibinfo{title}{{Changing the colour of light in a silicon
  resonator}}.
\newblock \emph{\bibinfo{journal}{Nature Photon.}}
  \textbf{\bibinfo{volume}{1}}, \bibinfo{pages}{293--296}
  (\bibinfo{year}{2007}).

\bibitem{Tanabe2009}
\bibinfo{author}{Tanabe, T.}, \bibinfo{author}{Notomi, M.},
  \bibinfo{author}{Taniyama, H.} \& \bibinfo{author}{Kuramochi, E.}
\newblock \bibinfo{title}{{Dynamic release of trapped light from an ultrahigh-Q
  nanocavity via adiabatic frequency tuning}}.
\newblock \emph{\bibinfo{journal}{Phys. Rev. Lett.}}
  \textbf{\bibinfo{volume}{102}}, \bibinfo{pages}{043907}
  (\bibinfo{year}{2009}).

\bibitem{Dong2008}
\bibinfo{author}{Dong, P.}, \bibinfo{author}{Preble, S.~F.},
  \bibinfo{author}{Robinson, J.~T.}, \bibinfo{author}{Manipatruni, S.} \&
  \bibinfo{author}{Lipson, M.}
\newblock \bibinfo{title}{{Inducing Photonic Transitions between Discrete Modes
  in a Silicon Optical Microcavity}}.
\newblock \emph{\bibinfo{journal}{Phys. Rev. Lett.}}
  \textbf{\bibinfo{volume}{100}}, \bibinfo{pages}{033904}
  (\bibinfo{year}{2008}).

\bibitem{Holloway2012}
\bibinfo{author}{Holloway, C.~L.} \emph{et~al.}
\newblock \bibinfo{title}{{An overview of the theory and applications of
  metasurfaces: The two-dimensional equivalents of metamaterials}}.
\newblock \emph{\bibinfo{journal}{IEEE Antennas Propag. Mag.}}
  \textbf{\bibinfo{volume}{54}}, \bibinfo{pages}{10--35}
  (\bibinfo{year}{2012}).

\bibitem{Kildishev2013}
\bibinfo{author}{Kildishev, A.~V.}, \bibinfo{author}{Boltasseva, A.} \&
  \bibinfo{author}{Shalaev, V.~M.}
\newblock \bibinfo{title}{{Planar Photonics with Metasurfaces}}.
\newblock \emph{\bibinfo{journal}{Science}} \textbf{\bibinfo{volume}{339}},
  \bibinfo{pages}{1232009} (\bibinfo{year}{2013}).

\bibitem{Yu2014a}
\bibinfo{author}{Yu, N.} \& \bibinfo{author}{Capasso, F.}
\newblock \bibinfo{title}{{Flat optics with designer metasurfaces.}}
\newblock \emph{\bibinfo{journal}{Nat. Mater.}} \textbf{\bibinfo{volume}{13}},
  \bibinfo{pages}{139--150} (\bibinfo{year}{2014}).

\bibitem{Li2017}
\bibinfo{author}{Li, G.}, \bibinfo{author}{Zhang, S.} \&
  \bibinfo{author}{Zentgraf, T.}
\newblock \bibinfo{title}{{Nonlinear photonic metasurfaces}}.
\newblock \emph{\bibinfo{journal}{Nat. Rev. Mater.}}
  \textbf{\bibinfo{volume}{2}}, \bibinfo{pages}{17010} (\bibinfo{year}{2017}).

\bibitem{Krasnok2017}
\bibinfo{author}{Krasnok, A.}, \bibinfo{author}{Tymchenko, M.} \&
  \bibinfo{author}{Alu, A.}
\newblock \bibinfo{title}{{Nonlinear Metasurfaces: A Paradigm Shift in
  Nonlinear Optics}}.
\newblock \emph{\bibinfo{journal}{Mater. Today}} \bibinfo{pages}{in print}
  (\bibinfo{year}{2017}).

\bibitem{Yu2011}
\bibinfo{author}{Yu, N.} \emph{et~al.}
\newblock \bibinfo{title}{{Light Propagation with Phase Discontinuities:
  Generalized Laws of Reflection and Refraction}}.
\newblock \emph{\bibinfo{journal}{Science}} \textbf{\bibinfo{volume}{334}},
  \bibinfo{pages}{333--337} (\bibinfo{year}{2011}).

\bibitem{Li2015}
\bibinfo{author}{Li, G.} \emph{et~al.}
\newblock \bibinfo{title}{{Continuous control of the nonlinearity phase for harmonic generations}}.
\newblock \emph{\bibinfo{journal}{Nature Mater.}} \textbf{\bibinfo{volume}{14}},
  \bibinfo{pages}{607--612} (\bibinfo{year}{2015}).


\bibitem{Emani2012}
\bibinfo{author}{Emani, N.~K.} \emph{et~al.}
\newblock \bibinfo{title}{{Electrically tunable damping of plasmonic resonances
  with graphene}}.
\newblock \emph{\bibinfo{journal}{Nano Lett.}} \textbf{\bibinfo{volume}{12}},
  \bibinfo{pages}{5202--5206} (\bibinfo{year}{2012}).

\bibitem{Yao2013}
\bibinfo{author}{Yao, Y.} \emph{et~al.}
\newblock \bibinfo{title}{{Broad electrical tuning of graphene-loaded plasmonic
  antennas}}.
\newblock \emph{\bibinfo{journal}{Nano Lett.}} \textbf{\bibinfo{volume}{13}},
  \bibinfo{pages}{1257--1264} (\bibinfo{year}{2013}).

\bibitem{Dabidian2015}
\bibinfo{author}{Dabidian, N.} \emph{et~al.}
\newblock \bibinfo{title}{{Electrical switching of infrared light using
  graphene integration with plasmonic Fano resonant metasurfaces}}.
\newblock \emph{\bibinfo{journal}{ACS Photon.}} \textbf{\bibinfo{volume}{2}},
  \bibinfo{pages}{216--227} (\bibinfo{year}{2015}).

\bibitem{Dabidian2016}
\bibinfo{author}{Dabidian, N.} \emph{et~al.}
\newblock \bibinfo{title}{{Experimental demonstration of phase modulation and
  motion sensing using graphene-integrated metasurfaces}}.
\newblock \emph{\bibinfo{journal}{Nano Lett.}} \textbf{\bibinfo{volume}{16}},
  \bibinfo{pages}{3607--3615} (\bibinfo{year}{2016}).

\bibitem{Klein2006}
\bibinfo{author}{Klein, M.~W.}, \bibinfo{author}{Enkrich, C.},
  \bibinfo{author}{Wegener, M.} \& \bibinfo{author}{Linden, S.}
\newblock \bibinfo{title}{{Second-harmonic generation from magnetic
  metamaterials.}}
\newblock \emph{\bibinfo{journal}{Science}} \textbf{\bibinfo{volume}{313}},
  \bibinfo{pages}{502--504} (\bibinfo{year}{2006}).

\bibitem{Lee2014}
\bibinfo{author}{Lee, J.} \emph{et~al.}
\newblock \bibinfo{title}{{Giant nonlinear response from plasmonic metasurfaces
  coupled to intersubband transitions.}}
\newblock \emph{\bibinfo{journal}{Nature}} \textbf{\bibinfo{volume}{511}},
  \bibinfo{pages}{65--69} (\bibinfo{year}{2014}).

\bibitem{Vampa2017}
\bibinfo{author}{Vampa, G.} \emph{et~al.}
\newblock \bibinfo{title}{{Plasmon-enhanced high-harmonic generation from
  silicon}}.
\newblock \emph{\bibinfo{journal}{Nat. Phys.}} \textbf{\bibinfo{volume}{13}},
  \bibinfo{pages}{659--662} (\bibinfo{year}{2017}).

\bibitem{Wurtz2011}
\bibinfo{author}{Wurtz, G.~A.} \emph{et~al.}
\newblock \bibinfo{title}{{Designed ultrafast optical nonlinearity in a
  plasmonic nanorod metamaterial enhanced by nonlocality}}.
\newblock \emph{\bibinfo{journal}{Nat. Nanotechnol.}}
  \textbf{\bibinfo{volume}{6}}, \bibinfo{pages}{106--110}
  (\bibinfo{year}{2011}).

\bibitem{Guo2016}
\bibinfo{author}{Guo, P.}, \bibinfo{author}{Schaller, R.~D.},
  \bibinfo{author}{Ketterson, J.~B.} \& \bibinfo{author}{Chang, R. P.~H.}
\newblock \bibinfo{title}{{Ultrafast switching of tunable infrared plasmons in
  indium tin oxide nanorod arrays with large absolute amplitude}}.
\newblock \emph{\bibinfo{journal}{Nat. Photon.}} \textbf{\bibinfo{volume}{10}},
  \bibinfo{pages}{267--273} (\bibinfo{year}{2016}).

\bibitem{Shcherbakov2017}
\bibinfo{author}{Shcherbakov, M.~R.} \emph{et~al.}
\newblock \bibinfo{title}{{Ultrafast all-optical tuning of direct-gap
  semiconductor metasurfaces}}.
\newblock \emph{\bibinfo{journal}{Nat. Commun.}} \textbf{\bibinfo{volume}{8}},
  \bibinfo{pages}{17} (\bibinfo{year}{2017}).

\bibitem{Kuznetsov2016a}
\bibinfo{author}{Kuznetsov, A.~I.}, \bibinfo{author}{Miroshnichenko, A.~E.},
  \bibinfo{author}{Brongersma, M.~L.}, \bibinfo{author}{Kivshar, Y.~S.} \&
  \bibinfo{author}{Lukyanchuk, B.}
\newblock \bibinfo{title}{{Optically resonant dielectric nanostructures}}.
\newblock \emph{\bibinfo{journal}{Science}} \textbf{\bibinfo{volume}{354}},
  \bibinfo{pages}{aag2472} (\bibinfo{year}{2016}).

\bibitem{Wu2014}
\bibinfo{author}{Wu, C.} \emph{et~al.}
\newblock \bibinfo{title}{{Spectrally selective chiral silicon metasurfaces
  based on infrared Fano resonances.}}
\newblock \emph{\bibinfo{journal}{Nat. Commun.}} \textbf{\bibinfo{volume}{5}},
  \bibinfo{pages}{3892} (\bibinfo{year}{2014}).

\bibitem{Yang2014a}
\bibinfo{author}{Yang, Y.}, \bibinfo{author}{Kravchenko, I.~I.},
  \bibinfo{author}{Briggs, D.~P.} \& \bibinfo{author}{Valentine, J.}
\newblock \bibinfo{title}{{All-dielectric metasurface analogue of
  electromagnetically induced transparency}}.
\newblock \emph{\bibinfo{journal}{Nat. Commun.}} \textbf{\bibinfo{volume}{5}},
  \bibinfo{pages}{5753} (\bibinfo{year}{2014}).

\bibitem{Shcherbakov2014b}
\bibinfo{author}{Shcherbakov, M.~R.} \emph{et~al.}
\newblock \bibinfo{title}{{Enhanced Third-Harmonic Generation in Silicon
  Nanoparticles Driven by Magnetic Response}}.
\newblock \emph{\bibinfo{journal}{Nano Lett.}} \textbf{\bibinfo{volume}{14}},
  \bibinfo{pages}{6488--6492} (\bibinfo{year}{2014}).

\bibitem{Liu2016c}
\bibinfo{author}{Liu, S.} \emph{et~al.}
\newblock \bibinfo{title}{{Resonantly Enhanced Second-Harmonic Generation Using
  III--V Semiconductor All-Dielectric Metasurfaces}}.
\newblock \emph{\bibinfo{journal}{Nano Lett.}} \textbf{\bibinfo{volume}{16}},
  \bibinfo{pages}{5426--5432} (\bibinfo{year}{2016}).

\bibitem{Grinblat2016}
\bibinfo{author}{Grinblat, G.}, \bibinfo{author}{Li, Y.},
  \bibinfo{author}{Nielsen, M.~P.}, \bibinfo{author}{Oulton, R.~F.} \&
  \bibinfo{author}{Maier, S.~A.}
\newblock \bibinfo{title}{{Efficient Third Harmonic Generation and Nonlinear
  Sub-Wavelength Imaging at a Higher-Order Anapole Mode in a Single Germanium
  Nanodisk}}.
\newblock \emph{\bibinfo{journal}{ACS Nano}} \textbf{\bibinfo{volume}{11}},
  \bibinfo{pages}{953--960} (\bibinfo{year}{2016}).

\bibitem{Makarov2017}
\bibinfo{author}{Makarov, S.~V.} \emph{et~al.}
\newblock \bibinfo{title}{{Efficient Second-Harmonic Generation in
  Nanocrystalline Silicon Nanoparticles}}.
\newblock \emph{\bibinfo{journal}{Nano Lett.}} \textbf{\bibinfo{volume}{17}},
  \bibinfo{pages}{3047--3053} (\bibinfo{year}{2017}).

\bibitem{Shalaev2015}
\bibinfo{author}{Shalaev, M.~I.} \emph{et~al.}
\newblock \bibinfo{title}{{High-Efficiency All-Dielectric Metasurfaces for
  Ultracompact Beam Manipulation in Transmission Mode}}.
\newblock \emph{\bibinfo{journal}{Nano Lett.}} \textbf{\bibinfo{volume}{15}},
  \bibinfo{pages}{6261--6266} (\bibinfo{year}{2015}).

\bibitem{Neuner2013}
\bibinfo{author}{Neuner, B.} \emph{et~al.}
\newblock \bibinfo{title}{{Efficient infrared thermal emitters based on
  low-albedo polaritonic meta-surfaces}}.
\newblock \emph{\bibinfo{journal}{Appl. Phys. Lett.}}
  \textbf{\bibinfo{volume}{102}}, \bibinfo{pages}{211111}
  (\bibinfo{year}{2013}).

\bibitem{Drozdov2012}
\bibinfo{author}{Drozdov, A. A.} \emph{et~al.}
\newblock \bibinfo{title}{{Self-phase modulation and frequency generation with few-cycle optical pulses in nonlinear dispersive media}}.
\newblock \emph{\bibinfo{journal}{Phys. Rev. A}}
  \textbf{\bibinfo{volume}{86}}, \bibinfo{pages}{053822}
  (\bibinfo{year}{2012}).


\bibitem{Buyanovskaya2017}
\bibinfo{author}{Buyanovskaya, E. M.} \emph{et~al.}
\newblock \bibinfo{title}{{Harmonic generation enhancement due to interaction of few-cycle light pulses in nonlinear dielectric coating on a mirror}}.
\newblock \emph{\bibinfo{journal}{Phys. Lett. A}}
  \textbf{\bibinfo{volume}{381}}, \bibinfo{pages}{3714--3721}
  (\bibinfo{year}{2017}).




\bibitem{Shorokhov2016a}
\bibinfo{author}{Shorokhov, A.~S.} \emph{et~al.}
\newblock \bibinfo{title}{{Multifold Enhancement of Third-Harmonic Generation
  in Dielectric Nanoparticles Driven by Magnetic Fano Resonances}}.
\newblock \emph{\bibinfo{journal}{Nano Lett.}} \textbf{\bibinfo{volume}{16}},
  \bibinfo{pages}{4857--4861} (\bibinfo{year}{2016}).



\bibitem{Zhang2011}
\bibinfo{author}{Zhang, J.} \emph{et~al.}
\newblock \bibinfo{title}{{Saturation of the second harmonic generation from GaAs-filled metallic hole arrays by nonlinear absorption}}.
\newblock \emph{\bibinfo{journal}{Phys. Rev. B}} \textbf{\bibinfo{volume}{83}},
\bibinfo{pages}{165438} (\bibinfo{year}{2011}).


\bibitem{Zhang2012}
\bibinfo{author}{Zhang, J.} \emph{et~al.}
\newblock \bibinfo{title}{{Free carrier induced spectral shift for GaAs filled metallic hole arrays}}.
\newblock \emph{\bibinfo{journal}{Opt. Express}} \textbf{\bibinfo{volume}{20}},
\bibinfo{pages}{7142--7150} (\bibinfo{year}{2012}).




\bibitem{Haus1984}
\bibinfo{author}{Haus, H.~A.}
\newblock \emph{\bibinfo{title}{{Waves And Fields In Optoelectronics}}}
  (\bibinfo{publisher}{Prentice-Hall}, \bibinfo{address}{New Jersey},
  \bibinfo{year}{1984}).

\bibitem{minkov2017}
\bibinfo{author}{Minkov, M.}, \bibinfo{author}{Shi, Y.} \&
  \bibinfo{author}{Fan, S.}
\newblock \bibinfo{title}{{Exact solution to the steady-state dynamics of a
  periodically-modulated resonator}}.
\newblock \emph{\bibinfo{journal}{APL Photonics}} \textbf{\bibinfo{volume}{2}},
  \bibinfo{pages}{076101} (\bibinfo{year}{2017}).

\bibitem{Weiner2000}
\bibinfo{author}{Weiner, A.~M.}
\newblock \bibinfo{title}{{Femtosecond pulse shaping using spatial light
  modulators}}.
\newblock \emph{\bibinfo{journal}{Rev. Sci. Instrum.}}
  \textbf{\bibinfo{volume}{71}}, \bibinfo{pages}{1929} (\bibinfo{year}{2000}).

\bibitem{Neyra2016}
\bibinfo{author}{Neyra, E.} \emph{et~al.}
\newblock \bibinfo{title}{{Extending the high-order harmonic generation cutoff by means of self-phase-modulated chirped pulses}}.
\newblock \emph{\bibinfo{journal}{Laser Phys. Lett.}}
  \textbf{\bibinfo{volume}{13}}, \bibinfo{pages}{115303} (\bibinfo{year}{2016}).


\bibitem{Tsakmakidis2017}
\bibinfo{author}{Tsakmakidis, K.~L.} \emph{et~al.}
\newblock \bibinfo{title}{{Breaking Lorentz reciprocity to overcome the time-bandwidth limit in physics and engineering}}.
\newblock \emph{\bibinfo{journal}{Science}}
  \textbf{\bibinfo{volume}{356}}, \bibinfo{pages}{1260--1264} (\bibinfo{year}{2017}).

\bibitem{Grigorescu2009}
\bibinfo{author}{Grigorescu, A.~E.} \&  \bibinfo{author}{Hagen, C.~W.}
\newblock \bibinfo{title}{{Resists for sub-20-nm electron beam lithography with a focus on HSQ: state of the art}}.
\newblock \emph{\bibinfo{journal}{Nanotechnol.}}
  \textbf{\bibinfo{volume}{20}}, \bibinfo{pages}{292001} (\bibinfo{year}{2009}).

\bibitem{Bennett1990}
\bibinfo{author}{Bennett, B. R.}, \bibinfo{author}{Soref, R. A.}    \&  \bibinfo{author}{Del Alamo, J. A.}
\newblock \bibinfo{title}{{Carrier-induced change in refractive index of InP, GaAs, and InGaAsP}}.
\newblock \emph{\bibinfo{journal}{IEEE J. Quant. Electron.}}
  \textbf{\bibinfo{volume}{26}}, \bibinfo{pages}{113--122} (\bibinfo{year}{1990}).

\bibitem{Louisy2015}
\bibinfo{author}{Louisy, M.} \emph{et~al.}
\newblock \bibinfo{title}{{Gating attosecond pulses in a noncollinear geometry}}.
\newblock \emph{\bibinfo{journal}{Optica}}
\textbf{\bibinfo{volume}{2}}, \bibinfo{pages}{563--566} (\bibinfo{year}{2015}).

\bibitem{Sheik1988}
\bibinfo{author}{Sheik-Bahae, M.} \& \bibinfo{author}{Kwok, H.~S.}
\newblock \bibinfo{title}{{Controlled CO$_2$ laser melting of silicon}}.
\newblock \emph{\bibinfo{journal}{J. Appl. Phys.}}
  \textbf{\bibinfo{volume}{63}}, \bibinfo{pages}{518} (\bibinfo{year}{1988}).

\bibitem{Osamura1972}
\bibinfo{author}{Osamura, K.} \& \bibinfo{author}{Murakami, Y.}
\newblock \bibinfo{title}{{Free Carrier Absorption in n-GaAs}}.
\newblock \emph{\bibinfo{journal}{J. J. Appl. Phys.}}
  \textbf{\bibinfo{volume}{11}}, \bibinfo{pages}{365} (\bibinfo{year}{1972}).

\bibitem{You2017}
\bibinfo{author}{You, Y.~S.} \emph{et~al.}
\newblock \bibinfo{title}{{High-harmonic generation in amorphous solids}}.
\newblock \emph{\bibinfo{journal}{Nat. Commun.}}
  \textbf{\bibinfo{volume}{8}}, \bibinfo{pages}{724} (\bibinfo{year}{2017}).






\end{thebibliography}
\end{document}